\documentclass[]{article}
\usepackage[dvipsnames]{xcolor}
\usepackage{graphicx}

\usepackage[version=4]{mhchem}
\usepackage{amsmath}
\usepackage{amssymb}
\usepackage{color}
\usepackage{overcite}
\usepackage[font=small,labelfont=bf]{caption}

\usepackage{amsfonts}
\usepackage{bm}
\usepackage{bbm}
\usepackage{braket}
\usepackage{mathtools}
\usepackage{listings}
\usepackage[Symbol]{upgreek}
\usepackage{subcaption}
\usepackage{algorithm}
\usepackage{algpseudocode}
\usepackage{hyperref}
\hypersetup{
    colorlinks=true,
    linkcolor=blue,
    citecolor=red,
    filecolor=magenta,      
    urlcolor=cyan
}

\DeclareMathOperator{\el}{\mathrm{e}^{--}}
\DeclareMathOperator{\pr}{\mathrm{p}^{+}}
\DeclareMathOperator{\Tr}{\mathrm{Tr}}

\DeclareMathOperator{\Gr}{\mathrm{Gr}_{\textit{m},\textit{n}}}
\DeclareMathOperator{\TGr}{ \textit{T}_\mathbf{D}\Gr}

\title{\bf Second-order self-consistent field algorithms: from classical to quantum nuclei}
\date{December 7, 2022}
\author{\vspace{0.3cm}
Robin Feldmann, Alberto Baiardi, and Markus Reiher\footnote{Corresponding author; e-mail: markus.reiher@phys.chem.ethz.ch}\\
        \textit{ETH Z\"urich, Laboratorium f\"ur Physikalische Chemie,}\\
        \textit{Vladimir-Prelog-Weg 2, 8093 Z\"urich, Switzerland}\\[2ex]
}

\begin{document}

\maketitle

\begin{abstract}
This work presents a general framework for deriving exact and approximate Newton self-consistent field (SCF) orbital optimization algorithms by leveraging concepts borrowed from differential geometry.
Within this framework, we extend the augmented Roothaan--Hall (ARH) algorithm to unrestricted electronic and nuclear-electronic calculations. We demonstrate that ARH yields an excellent compromise between stability and computational cost for SCF problems that are hard to converge with conventional first-order optimization strategies.
In the electronic case, we show that ARH overcomes the slow convergence of orbitals in strongly-correlated molecules with the example of several iron-sulfur clusters.
For nuclear-electronic calculations, ARH significantly enhances the convergence already for small molecules, as demonstrated for a series of protonated water clusters.
\end{abstract}

\newpage
\section{Introduction}
\label{sec:intro}

Processes that are strongly influenced by nuclear quantum effects are among the most daunting targets of quantum chemical simulations.
One often invokes the Born--Oppenheimer (BO) approximation\cite{Born-Oppenheimer1927,BornHuang1954,tully2000perspective} to adiabatically separate the electronic and the nuclear motion.
This separation becomes inaccurate as soon as the coupling between the electrons and nuclei is strong.
This inaccuracy can be cured by including non-adiabatic effects \textit{a posteriori}, e.g., with perturbation theory.\cite{Pachucki2008_NonadiabaticPerturbationTheory}
Alternatively, the full molecular Schr\"{o}dinger equation can be solved with so-called nuclear-electronic methods, treating nuclei and electrons on the same footing.
In this way, non-adiabatic and nuclear quantum effects are automatically included.

Explicitly correlated methods yield the most accurate solution of the full molecular Schr\"odinger equation.
They are, however, limited to few-particle molecules due to the factorial scaling of their computational cost with system size.\cite{Varga1995,Adamowicz2012_Review,Matyus2012_ECG1,Matyus2019_Review}
A cost-effective alternative is offered by orbital-based nuclear-electronic approaches.\cite{Thomas1969_protonic1,Thomas1969_protonic2,Thomas1970_protonic3,Thomas1971_protonic4,Nakai1998_NOMO-Original,Hammes-Schiffer2020_Review}
Although these methods usually do not reach the accuracy of their explicitly correlated counterparts, they can be routinely applied to molecules with dozens of electrons and multiple quantum nuclei.\cite{Hammes-Schiffer2021_MulticomponenDF}
Orbital-based nuclear-electronic methods have often been devised by extending algorithms originally designed for electronic problems to nuclear-electronic ones.\cite{Hammes-Schiffer2020_Review}
This has led to the emergence of the nuclear-electronic counterparts of many wave function-\cite{Pavosevic_OOMP2-Multicomponent,Pavosevic2019_MulticomponentCC,Pavovsevic2019_NeoEomCCSD,pavovsevic2019_multicomponentCC2,Brorsen2020_SelectedCI-PreBO,Brorsen2021_multicomponentCASSCF,Muolo2020_Nuclear,Fajen2020_Separation,Pavosevic2021_MulticomponentUCC,Fajen2021_MulticomponentMP4,Brorsen2021_HBCIPT,Feldmann2022_QuantumProton} and density functional theory-based\cite{Hammes-Schiffer2017_MulticomponentDFT,Brorsen2018_TransferableDFT,HammesSchiffer2019_GGA-DFT,Mejia2019_MulticomponentDftDf} electronic structure methods.
These developments have revealed that some concepts cannot be straightforwardly transferred from electronic problems to nuclear-electronic ones.
For instance, molecules with a weakly correlated electronic ground state may display a strongly-correlated nuclear-electronic wave function.\cite{Brorsen2020_SelectedCI-PreBO,Feldmann2022_QuantumProton}
Quantum entanglement measures extracted from nuclear-electronic density matrix renormalization group (DMRG) calculations can guide this classification but it was shown that in the nuclear- electronic case it is difficult to classify a molecule unambiguously as strongly or weakly correlated.\cite{Muolo2020_Nuclear,Feldmann2022_QuantumProton}
Consequently, the efficiency of an algorithm originally ideated for electronic problems may change drastically when applied to nuclear-electronic problems.

In this work, we address the challenge that optimizing nuclear-electronic orbitals with self-consistent field (SCF) methods is significantly more difficult than for their purely electronic counterparts.
The convergence behavior of the nuclear-electronic Hartree--Fock optimization algorithms has received little attention in the literature so far, with the exceptions of Refs.~\citenum{Valeev2004_ENMO} and \citenum{HammesSchiffer2022_NeoNewDiis}.
As in electronic-structure theory, the Roothaan--Hall (RH) diagonalization method with level-shifting,\cite{Saunders1973_ScfLevelShift} damping,\cite{Hartree1947_ScfDamping} and the direct inversion of the iterative subspace (DIIS),\cite{Pulay1980_DIIS,Scuseria2002_EDIIS} represent the state of the art for nuclear-electronic calculations.
Recently,\cite{HammesSchiffer2022_NeoNewDiis} a new DIIS variant was developed for nuclear-electronic methods that simultaneously optimizes the nuclear and electronic coefficients of the wave function.
This algorithm currently represents the most efficient nuclear-electronic orbital optimization scheme.
However, as first-order optimization methods, they can converge either slowly, possibly to saddle points, or even diverge.
In electronic structure theory, these issues are especially relevant for strongly correlated systems.\cite{Scuserua2012_ScfAccComparison}
They can be partially alleviated, e.g., by orbital steering.\cite{Vaucher2017_OrbitalSteering}

The deficiencies of the RH approach can be overcome with Newton methods which, however, require the construction of the orbital Hessian.
This step represents the main bottleneck of the orbital optimization algorithm.\cite{Werner1980_Quadratically1,Werner1981_Quadratically2,Bacskay1981_OriginalQuadraticScf,Helgaker2014_Molecular,Saether2017_DensityBasedMLHF,Chan2017_CASSCF,Ma2017_DmrgScf,Werner2019_Quadratically3,Werner2019_Quadratically4,Helmich2021_OrcaTRAH,Umrigar2021_sCI-OrbitalOptimization}
For restricted Hartree--Fock calculations, this computational bottleneck can be overcome with the augmented Roothaan--Hall (ARH) method.\cite{Host2008_ArhCommunication,Host2008_ARH}
ARH is a quasi-Newton approximate second-order method that iteratively constructs the Hessian by expanding it in the basis of density matrices obtained in previous iteration steps. 

Here, we introduce a generic framework for Newton and quasi-Newton optimization algorithms for electronic and nuclear-electronic Hartree--Fock.
We derive the Newton equations by leveraging differential geometry techniques\cite{Absil2009_OptOnMatrixManifolds} and extend the ARH method to unrestricted-electronic and nuclear-electronic problems within this framework. 
We first demonstrate the efficiency of our new implementation for the unrestricted electronic case with the example of iron-sulfur clusters.
The electronic wave function of these systems is known to display strong correlation effects.\cite{Sharma2014_IronSulfur} 
This renders the solution of the SCF equations very challenging even with sophisticated convergence acceleration techniques.
Subsequently, we compare different optimization algorithms for nuclear-electronic Hartree--Fock calculations.
We choose as test cases water clusters because they are known to exhibit strong nuclear quantum effects.\cite{Kulig2013_WaterClusters,Fagiani2016_WaterClustersNQE,Richardson2016_SCIENCE,Bowman2019_Eigen,Hammes-Schiffer2021_MulticomponenDF,Richardson2021_WaterClustersChapter}
We compare the efficiency of our new ARH method with the geometric direct minimization (GDM)\cite{Voorhis2002geometric,Dunietz2002_GDM} and DIIS methods, as well as with a newly implemented exact Newton method.
We show that ARH is significantly faster than GDM and more robust than DIIS with a limited computational overhead compared to the much more demanding full Newton optimization.

The work is organized as follows:
We introduce in Sec.~\ref{sec:theory} the Newton method from a differential geometry perspective and show how to apply it to restricted Hartree--Fock theory.
We then extend the Newton method to unrestricted and nuclear-electronic problems and derive the corresponding exact Newton and ARH working equations in Sec.~\ref{sec:theory2}.
Afterwards, in Sec.~\ref{sec:trah} we review the trust-region augmented Hessian approach which we apply to ARH to ensure that the algorithm converges to a minimum of the energy functional.
After presenting the computational details in Sec.~\ref{sec:details}, we discuss the results for electronic and nuclear-electronic systems in Sec.~\ref{sec:results}.

\section{A differential geometry perspective on Hartree--Fock theory}
\label{sec:theory}

We start this section with a brief account of our formalism for the restricted electronic BO Hartree--Fock method.
We then describe the Hartree--Fock optimization from a differential geometry perspective and derive the Newton equations for the restricted Hartree--Fock method.
Although often overlooked, differential geometry is a powerful tool for studying optimization problems in quantum chemistry.\cite{Burton2021_EnergyLandscape,Lubich2015_TangetnSpaceDMRG,Baiardi2021_ElectronDynamics,Aoto2021_QCGrassmann}
Subsequently, we show how to enhance computational efficiency of the solution of the Newton equations.

\subsection{Restricted Hartree--Fock theory}

In Hartree--Fock theory, the $N$-electron Hartree--Fock wave function, $\Psi_\mathrm{BO}$, is taken as a Slater determinant, $\Phi$ defined as an antisymmetrized product of spin-orbitals, $\phi_{is}$,
\begin{equation}
  \Psi_\mathrm{BO} (\mathbf{r}) = \Phi (\mathbf{r}) 
    = \frac{1}{\sqrt{N!}} \mathcal{S}_- \left( \prod_s^S \prod^{N_s}_i \phi_{is}(\mathbf{r}_{is}) )\right).
  \label{eq:ansatzBO}
\end{equation}
Here, $\mathbf{r}$ is a vector collecting the positions of all electrons, $\mathbf{r}_{is}$ is the coordinate of the $i$-th electron with spin quantum number $S$ and spin projection onto the $z$-axis $s=-S, -S+1,\dots, S$. $N_{s}$ is the number of electrons with a given spin.
Although for electrons the total spin is $S=\frac{1}{2}$, we derive the equations for a generic $S$ to facilitate their extension to the nuclear-electronic case.
The antisymmetrization operator, $\mathcal{S}_-$, enforces the correct permutational symmetry of the wave function. 
The spin-orbitals $\phi_{is}$ are expressed as a linear combination of $N_{\mathrm{AO}}$ pre-defined Gaussian-type orbitals $\left\{ \chi_\mu(\mathbf{r}) \right\}_{\mu=1}^{N_\mathrm{AO}}$, which are referred to as atomic orbitals,
\begin{equation}
  \phi_{is}(\mathbf{r}_{is}) = \sum_{\mu=1}^{N_{\mathrm{AO}}} C_{s,i\mu} \chi_{\mu}(\mathbf{r}_{is}) \, .
  \label{eq:AOtoMO}
\end{equation}
The coefficients $C_{s,i\mu}$ of the orbitals with spin $s$ can be grouped into a matrix, $\mathbf{C}_{s}$, which can be further partitioned into an occupied block, $\mathbf{C}_{s,\mathrm{o}}$, of dimension $N_{\mathrm{AO}}\times N_{s}$, and a virtual block, $\mathbf{C}_{s,\mathrm{v}}$, of dimension $N_{\mathrm{AO}}\times( N_{\mathrm{AO}} - N_{s})$.
Only the occupied orbitals enter the Hartree--Fock wave function.
Therefore, its corresponding energy is completely determined by the AO density matrices, $\mathbf{D}_{s}$, which are defined as
\begin{equation}
  \mathbf{D}_{s} = \mathbf{C}_{s,\mathrm{o}} \mathbf{C}_{s,\mathrm{o}}^\mathrm{T}.
  \label{eq:densitymatrixUhf}
\end{equation}
The density matrices satisfy three conditions, namely idempotency, symmetry with respect to transposition, and the trace being equal to $N_s$:
\begin{equation}
  \mathbf{D}_{s}^2=\mathbf{D}_{s}=\mathbf{D}^\mathrm{T}_{s} ,\qquad \mathrm{Tr}\left(\mathbf{D}_{s}\right) = N_{s}.
  \label{eq:projectorProps}
\end{equation}
The electronic Hartree--Fock energy can be expressed as
\begin{equation}
    E_\mathrm{BO}(\{\mathbf{D}_{s}\}) = \sum_s^{S} \sum_{\mu\nu}^{N_{\mathrm{AO}}} h_{\mu\nu} D_{s\nu\mu}
      + \frac{1}{2} \sum_s^{S} \sum_{\mu\nu\rho\sigma}^{N_{\mathrm{AO}}} D_{s\nu\mu} \Bar{V}_{\mu\nu,\rho\sigma} D_{s\sigma\rho}
      + \sum_{s<t}^{S} \sum_{\mu\nu\rho\sigma}^{N_{\mathrm{AO}}} D_{s\nu\mu} V_{\mu\nu,\rho\sigma} D_{t\sigma\rho} \, ,
 \label{eq:UhfElectronicEnergy}
\end{equation}
where $h_{\mu\nu}$ contains the one-electron integrals, $V_{\mu\nu,\rho\sigma}$ the Coulomb integrals for basis functions $\mu,\nu$ of the first electron and $\rho,\sigma$ of the second one, and $\Bar{V}_{\mu\nu,\rho\sigma}$ denotes the antisymmetrized Coulomb integral.
Electrons are spin-$\frac{1}{2}$ Fermions for which the allowed values of the spin projection $s$ are $\frac{1}{2}$ and $-\frac{1}{2}$, which we will denote as $\uparrow$ and $\downarrow$, respectively.
In the restricted electronic Hartree--Fock method, the spatial part of the spin-orbitals are equal, i.e., $\mathbf{C}_{\uparrow} = \mathbf{C}_{\downarrow} = \mathbf{C}$ and, therefore, $\mathbf{D}_{\uparrow} = \mathbf{D}_{\downarrow} = \mathbf{D}$.
This reduces the number of free parameters and simplifies Eq.~(\ref{eq:UhfElectronicEnergy}) to
\begin{equation}
 \begin{aligned}
    E_\mathrm{BO}(\mathbf{D}) =\ &2\sum_{\mu\nu}^{N_{\mathrm{AO}}} h_{\mu\nu} D_{\nu\mu} + \frac12 \sum_{\mu\nu\rho\sigma}^{N_{\mathrm{AO}}} D_{\nu\mu} (2V_{\mu\nu,\rho\sigma}-V_{\mu\sigma,\rho\nu}) D_{\sigma\rho}.
 \end{aligned}
 \label{eq:RhfElectronicEnergy}
\end{equation}
The restricted Hartree--Fock method optimizes the energy, Eq.~(\ref{eq:RhfElectronicEnergy}), with respect to the orbital coefficients under the constraint that the orbitals remain orthonormal
\begin{equation}
  \mathrm{min} \Big\{ E(\mathbf{C}_{\mathrm{o}})\ \Big|\
                      \mathbf{C}_{\mathrm{o}} \in \mathbb{R}^{N_{\mathrm{AO}}\times N/2},\
                      \mathbf{C}_{\mathrm{o}}^\mathrm{T}\mathbf{C}_{\mathrm{o}}=\mathbbm{1}
               \Big\}.
  \label{eq:BO_EnergyFunctional}
\end{equation}
where $\mathbf{C}_\mathrm{o}$ is the occupied block of $\mathbf{C}$.
The energy minimization can be equivalently carried out with respect to the density matrix $\mathbf{D}$.
In this case, the orthogonality-constrained optimization problem reads
\begin{equation}
  \mathrm{min} \Big\{ E(\mathbf{D})\ \Big|\
                      \mathbf{D} \in \mathbb{R}^{N_{\mathrm{AO}}\times N_\mathrm{AO}},\ 
                      \mathbf{D}^2=\mathbf{D}=\mathbf{D}^\mathrm{T} ,\ \mathrm{Tr}\left(\mathbf{D}\right) = N/2.
               \Big\}.
  \label{eq:EnergyMinimization}
\end{equation}
In the following, we demonstrate how to solve these minimization problems by leveraging differential geometry. 
This mathematical discipline provides an elegant and convenient framework to derive efficient optimization algorithms for functions of matrix arguments with constraints on the parameters.\cite{Absil2009_OptOnMatrixManifolds}

\subsection{Optimizing the energy functional on a manifold}

The Newton method is the standard second-order method to optimize iteratively a real-valued function $f(\mathbf{x})$ with $\mathbf{x}\in \mathbb{R}^n$, that is at least twice differentiable.
Starting from an initial guess $\mathbf{x}_0$, the function is expanded at the $i$-th iteration around the corresponding approximate solution, $\mathbf{x}_i$, as
\begin{equation}
  f(\mathbf{x}_i+\mathbf{h}) = f(\mathbf{x}_i) 
                              + \mathrm{grad}_f(\mathbf{x}_i)^\mathrm{T}\mathbf{h} 
                              + \frac{1}{2}  \mathbf{h}^\mathrm{T}\mathrm{Hess}_f(\mathbf{x}_i)\mathbf{h}
                              + \mathcal{O}(\mathbf{h}^3).
    \label{eq:Taylorfx}
\end{equation}
Here, $\mathrm{Hess}_f(\mathbf{x}_i) $ and $\mathrm{grad}_f (\mathbf{x}_i) $ refer to the Hessian and gradient of $f$ at point $\mathbf{x}_i$, respectively.
Setting the gradient of Eq.~(\ref{eq:Taylorfx}) with respect to $\mathbf{h}$ to zero yields the Newton equation:
\begin{equation}
  \mathrm{Hess}_f(\mathbf{x}_i) \mathbf{h} = -\mathrm{grad}_f(\mathbf{x}_i) \, .
  \label{eq:NewtondEquation}
\end{equation}
The updated approximate solution $\mathbf{x}_{i+1}$ is obtained from the solution of Eq.~(\ref{eq:NewtondEquation}) as
\begin{equation}
  \mathbf{x}_{i+1} = \mathbf{x}_i + \mathbf{h}.
  \label{eq:updateXinRn}
\end{equation}
This algorithm converges to the stationary point closest to the starting guess $\mathbf{x}_0$.\cite{Absil2009_OptOnMatrixManifolds}

Naively applying the Newton method to Eq.~(\ref{eq:EnergyMinimization}) by updating $\mathbf{D}_i$ based on Eq.~(\ref{eq:updateXinRn}) would yield an updated density matrix, $\mathbf{D}_{i+1}$, that no longer fulfills the constraints of Eq.~(\ref{eq:projectorProps}).
The reason is that the density matrix is not defined on Euclidean space where the shortest path between two points is the difference between their coordinates. 
This renders the conventional rules from calculus on vector spaces, as well as the linear rule to update $\mathbf{D}_{i+1}$ as in Eq.~(\ref{eq:updateXinRn}), inapplicable.

Differential geometry studies manifolds, i.e., mathematical spaces that locally --- but not necessarily globally --- resemble Euclidean space.
In the case of the Hartree--Fock method, the energy $E(\mathbf{D})$ is defined on the manifold spanned by the matrices fulfilling the constraints of Eq.~(\ref{eq:projectorProps}). 
This manifold is called the Grassmann manifold $\Gr$ that is defined as\cite{Absil2020_GrassmannHandbook, helmke2007_NewtonGrassmann,cances2021convergence}
\begin{equation}
  \Gr = \left\{ \mathbf{D}\in\mathbb{R}^{n\times n}\, |\, 
                \mathbf{D}^2=\mathbf{D}=\mathbf{D}^\mathrm{T} ,\ \mathrm{Tr}\left(\mathbf{D}\right) = m
        \right\}.
  \label{eq:Grassmann}
\end{equation}
We show in the following how to evaluate gradients and Hessians, and consequently, how to define a Newton step on the Grassmann manifold.
First, we introduce two vector spaces --- i.e., spaces where the Euclidean differentiation rules apply --- which are the space of symmetric, $\mathrm{Sym}_n$, and antisymmetric, $\mathfrak{so}_n$, $n\times n$ matrices
\begin{align}
  \mathrm{Sym}_n &= \left\{ \mathbf{S} \in \mathbb{R}^{n\times n}\, |\, \mathbf{S}=\mathbf{S}^\mathrm{T} \right\}, 
  \label{eq:SymetricMatrices} \\
  \mathfrak{so}_n &= \left\{ \bm{\Omega} \in \mathbb{R}^{n\times n}\, |\, \bm{\Omega}^\mathrm{T}=-\bm{\Omega} \right\} \, .
  \label{eq:AntisymmetricMatrices}
\end{align}
The energy function $E$ defined in Eq.~(\ref{eq:RhfElectronicEnergy}) is a map from the vector space of symmetric matrices to the real numbers,
\begin{equation}
  E: \mathrm{Sym}_n \rightarrow \mathbb{R} \, .
  \label{eq:EnergySymmetric}
\end{equation}
However, only $E$ values obtained from density matrices belonging to the Grassmann manifold correspond to Hartree--Fock wave functions.
Hence, we introduce the restriction of the energy to the manifold, $ \mathcal{E} =E|_{\Gr}$, as a map from $\Gr$ to $\mathbb{R}$, as
\begin{equation}
  \mathcal{E} =E|_{\Gr}: \Gr \rightarrow \mathbb{R}.
  \label{eq:EnergyGrassman}
\end{equation}
To calculate the gradient and Hessian of $\mathcal{E}$, we introduce the tangent space, denoted by $T_\mathbf{D}\mathrm{Gr}_{m,n}$.
The tangent space is the vector space that corresponds to the linear approximation of $\Gr$ at a point $\mathbf{D}$. 
This means, in practice, that for some sufficiently small $\epsilon$, $\mathbf{D}^\prime=\mathbf{D}+\epsilon\bm{\xi}$ with $\bm{\xi}\in\TGr$, is still an element of $\Gr$.
The tangent space of $\Gr$ at $\mathbf{D}$ is given by\cite{helmke2007_NewtonGrassmann}
\begin{equation}
  \TGr = \left\{ [\mathbf{D}, \bm{\Omega}] \, |\, \bm{\Omega}\in \mathfrak{so}_n \right\}
       = \left\{ [\mathbf{D}, [\mathbf{D}, \mathbf{S}]] \, |\, \mathbf{S}\in \mathrm{Sym}_n \right\},
  \label{eq:TangentSpace}
\end{equation}
where $[\cdot,\cdot]$ is the commutator operator.
It can be verified up to first order in $\epsilon$ that $\mathbf{D}^\prime=\mathbf{D}+\epsilon[\mathbf{D}, \bm{\Omega}]$ is still an element of the manifold.
Since the tangent space is a vector space, the conventional differentiation rules apply.\cite{Absil2009_OptOnMatrixManifolds}
The gradients and Hessians of $\mathcal{E}$ can be calculated by projecting \textit{a posteriori} the derivatives of $E$ onto the tangent space $T_\mathbf{D}\mathrm{Gr}_{m,n}$.
To that end, we introduce the adjoint representation at $\mathbf{D}$, $\mathrm{ad}_\mathbf{D}$, as
\begin{equation}
  \mathrm{ad}_\mathbf{D}(\mathbf{X}) = [\mathbf{D}, \mathbf{X}],\quad \mathbf{X} \in \mathbb{R}^{n\times n} \, .
  \label{eq:AdjointRepresentation}
\end{equation}
It follows from Eq.~(\ref{eq:TangentSpace}) that $\mathrm{ad}_\mathbf{D}$ maps any antisymmetric matrix $\bm{\Omega}$ onto $\TGr$
\begin{equation}
  \mathrm{ad}_\mathbf{D}(\bm{\Omega}) = [\mathbf{D},\bm{\Omega}] \ \in \TGr \, ,
  \label{eq:AdjointRepresentation_Antisymmetric}
\end{equation}
In the case of a symmetric matrix, the squared operator, $\pi_\mathbf{D}=\mathrm{ad}^2_\mathbf{D}$, acts as a projection\cite{helmke2007_NewtonGrassmann} onto $\TGr$ as
\begin{equation}
  \pi_\mathbf{D}(\mathbf{S}) = \mathrm{ad}^2_\mathbf{D}(\mathbf{S}) = [\mathbf{D},[\mathbf{D},\mathbf{S}]] \ \in \TGr \, .
  \label{eq:TangentSpaceProjection}
\end{equation}
This enables us to evaluate the gradient and Hessian of $\mathcal{E}$.
Since $E$ is defined on the vector space $\mathrm{Sym}_n$ we can differentiate $E$ twice with respect to $\mathbf{D}$ by applying the conventional differentiation rules.
The Fock matrix in this work is in all cases defined as the gradient of $E$ 
\begin{equation*}
  \mathbf{F}(\mathbf{D})=\mathrm{grad}_E(\mathbf{D)}\in \mathrm{Sym}_n.
  \label{eq:FockMatrixDefinition}
\end{equation*}
Note that, for the restricted case, the gradient of the energy $E$ with respect to the density matrix $\mathbf{D}$ contains a factor of 2 that arises from the assumption that $\mathbf{D}_{\uparrow} = \mathbf{D}_{\downarrow}$.
We adopt this definition of the Fock matrix because it enables us to derive the same working equations for the restricted and unrestricted cases.
We define the elements of the Hessian matrix of $E$ as
\begin{equation}
  E^{\mu\nu,\sigma\rho} = \frac{\partial^2 E}{\partial D_{\mu\nu} \partial D_{\sigma\rho}} 
                        = \left( \frac{\partial \mathbf{F}}{\partial D_{\mu\nu}} \right)_{\sigma\rho} \, .
  \label{eq:SecondDerivative}
\end{equation}
In differential geometry, the Hessian is defined by its action on a generic symmetric matrix, $\mathbf{S}$, as a linear operator, $\mathbf{H}_{E} (\mathbf{D}):\mathrm{Sym}_n \rightarrow \mathrm{Sym}_n$, as
\begin{equation}
  \mathbf{H}_E(\mathbf{D})(\mathbf{S}) = \sum_{\mu\nu}^{n} \frac{\partial \mathbf{F}}{\partial D_{\mu\nu}} S_{\mu\nu},
  \label{eq:HessianAction}
\end{equation}
where $\mathbf{D}$ is the point on the manifold at which the Hessian is evaluated.
The gradient and Hessian of a function on $\Gr$ were derived in Refs.~\citenum{helmke2007_NewtonGrassmann} and \citenum{absil2013GrassmannHessian2}.
The gradient is given by the projection of the Fock matrix onto the tangent space
\begin{equation}
  \mathbf{g}_\mathcal{E}(\mathbf{D}) 
    = \mathrm{grad}_\mathcal{E} (\mathbf{D} ) =
     \pi_\mathbf{D}(\mathbf{F})
    = [\mathbf{D},[\mathbf{D},\mathbf{F}]].
  \label{eq:GradientManifold}
\end{equation}
The Hessian is defined as a linear operator, $\mathbf{H}_\mathcal{E}(\mathbf{D)}:\TGr\rightarrow\TGr$ according to 
\begin{equation}
 \begin{aligned}
  \mathbf{H}_\mathcal{E} (\mathbf{D} )(\bm{\xi} ) 
    & = \pi_\mathbf{D} \big( \mathbf{H}_E (\mathbf{D})(\bm{\xi} )  
                             - \mathrm{ad}_\mathbf{D} \mathrm{ad}_{\mathbf{F}} (\bm{\xi})
                       \big) \\
    & = \mathrm{ad}^2_\mathbf{D} \big( \mathbf{H}_E (\mathbf{D})(\bm{\xi}) \big)
      - \mathrm{ad}_\mathbf{D} \mathrm{ad}_{\mathbf{F}} (\bm{\xi}) \, ,
 \end{aligned}
 \label{eq:HessianManifold}
\end{equation}
where we have exploited the following property to simplify the second term\cite{helmke2007_NewtonGrassmann}
\begin{equation}
  \mathrm{ad}^3_{\mathbf{D}}(\mathbf{X)}=\mathrm{ad}_{\mathbf{D}}(\mathbf{X)} \quad\mathrm{for}\ \mathbf{X} \in \mathbb{R}^{n\times n} \, .
  \label{eq:SimplificationAdjointRepresentation}
\end{equation}
Consequently, the Newton equations on the Grassmann manifold can be expressed compactly as
\begin{equation}
  \mathbf{H}_\mathcal{E}(\mathbf{D})\big(\bm{\xi}) = - \mathbf{g}_\mathcal{E}(\mathbf{D}).
  \label{eq:NewtonRhf}
\end{equation}
The Newton equations can be solved for the vector $\bm{\xi}$, which is an element of the tangent space and gives the step in the direction of the next stationary point.
From Eq.~(\ref{eq:TangentSpace}) it follows that $\bm{\xi}$ can be parametrized as 
\begin{equation}
  \bm{\xi}=\mathrm{ad}_\mathbf{D}(\bm{\Omega}), \quad \mathrm{with}\  \bm{\Omega}\in\mathfrak{so}_n.
\end{equation}
In practice, Eq.~(\ref{eq:NewtonRhf}) can be solved for $\bm{\Omega}$ by any direct solver for linear systems of equations.
To find the rule to update $\mathbf{D}$, given $\bm{\Omega}$, we need to find the analogue of Eq.~(\ref{eq:updateXinRn}) on $\Gr$. 
In Euclidean space, the solution vector, $\mathbf{x}_i$, is updated by propagating it in the direction parallel to the Newton vector, $\mathbf{h}$, see Eq.~(\ref{eq:updateXinRn}).
As mentioned above, such a procedure cannot be applied to a manifold because the density matrix would leave the manifold if propagated along the direction parallel to $\bm{\xi}$.
In order to consistently define an update rule, the concept of a straight line must be extended to manifolds.
In Euclidean space, a straight line $\mathbf{x}(t)$ with $t\in\mathbb{R}$, is the curve with zero acceleration for all $t$, i.e., $\Ddot{\mathbf{x}}(t)=0$.
The solution of this trivial differential equation with the initial conditions $\mathbf{x}(0)=\mathbf{x}_i$ and $\Dot{\mathbf{x}}(0)=\mathbf{h}$ is $\mathbf{x}(t)=\mathbf{x}_i+t\mathbf{h}$.
This idea can be generalized to define the unique curve, $\mathbf{D}(t)\in\Gr$, with initial conditions $\mathbf{D}(0)=\mathbf{D}_i$ and that coincides with the tangent vector $\bm{\xi}_i$ for $t=0$.
This curve is called geodesic and is the generalization of a straight line on a manifold.
As shown in Ref.~\citenum{helmke2007_NewtonGrassmann}, this curve is obtained on $\Gr$ as
\begin{equation}
  \mathbf{D}(t) = \mathrm{e}^{-t( \mathrm{ad}_{\mathbf{D}_i}(\bm{\xi}_i))} \ 
                  \mathbf{D}_i\ 
                  \mathrm{e}^{t( \mathrm{ad}_{\mathbf{D}_i}(\bm{\xi}_i))}.
  \label{eq:geodesic}
\end{equation}
\begin{figure}[htbp!]
  \centering
  \includegraphics[width=0.6\textwidth]{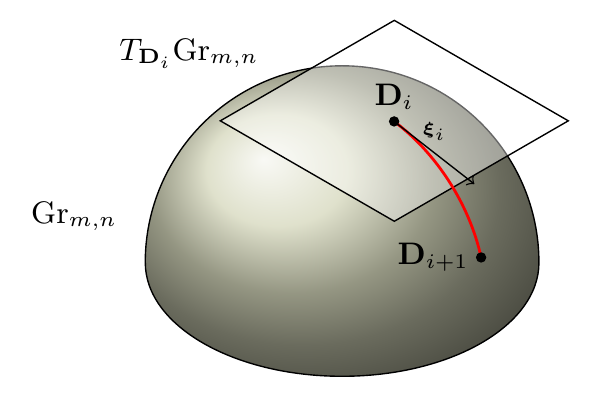}
  \caption{Schematic representation of the update of the density matrix on the Grassmann manifold.
  Depicted is the manifold and the tangent plane at the $i$-th iteration.
  The tangent vector $\bm{\xi}_i$ points in the direction of the next density matrix.
  The red curve is the geodesic.}
  \label{fig:manifold}
\end{figure}
We illustrate this concept in Figure~\ref{fig:manifold} with the example of a sphere embedded in three-dimensional space.
The manifold is represented by the sphere and the tangent space at the current iteration is depicted as the tangent plane at the point $\mathbf{D}_i$ on the manifold.
Solving the Newton equation yields the tangent vector $\bm{\xi}_i$ that gives the direction towards the next density matrix $\mathbf{D}_{i+1}$ that is closer to the next stationary point.
The update of the density matrix on $\Gr$ is obtained with Eq.~(\ref{eq:geodesic}) for $\bm{\xi}_i=\mathrm{ad}_\mathbf{D}(\bm{\Omega}_i)$ as 
\begin{equation}
  \mathbf{D}_{i+1} = \mathrm{e}^{-\mathrm{ad}^2_{\mathbf{D}_i} (\bm{\Omega})} \ 
                     \mathbf{D}_i \ 
                     \mathrm{e}^{\mathrm{ad}^2_{\mathbf{D}_i}(\bm{\Omega})},
  \label{eq:ExponentialMap}
\end{equation}
which is the analog of Eq.~(\ref{eq:updateXinRn}) for $\Gr$.
We obtain Eq.~(\ref{eq:ExponentialMap}) by setting $t=1$ and will discuss more refined strategies to adjust the step length in a later section.

\subsection{Cost-effective solution to the Newton equations}

In this section, we derive an efficient way to solve Eq.~(\ref{eq:NewtonRhf}) based on two observations:
First, Eq.~(\ref{eq:NewtonRhf}) is formulated in the space of $n \times n$ matrices, but the dimension of $\Gr$ is $m(n-m)$.
The energy depends on $N_\mathrm{AO}N_\mathrm{occ}$ coefficients, but it is invariant under unitary transformations of the occupied-occupied block of size $N^2_\mathrm{occ}$.
Hence, the dimension is $N_\mathrm{AO}N_\mathrm{occ}-N^2_\mathrm{occ}=N_\mathrm{occ}(N_\mathrm{AO}-N_\mathrm{occ})$.
This reflects the fact that only rotations between virtual and occupied orbitals change the Hartree--Fock energy. 
Second, the update of the density matrix of Eq.~(\ref{eq:ExponentialMap}) is given in terms of an exponential of a possibly very large matrix, which is computationally demanding to evaluate.
Hence, it would be advantageous to find a more efficient way to update the density matrix.

First, we simplify the Newton equations by realizing that the left- and right-hand side of the Newton equations, Eqs.~(\ref{eq:GradientManifold}) and (\ref{eq:HessianManifold}), respectively, are of the form $\pi_\mathbf{D}(\cdot)=\mathrm{ad}_\mathbf{D}^2(\cdot)$. 
With the following identity\cite{cances2021convergence}
\begin{equation}
    \bm{\xi} = \mathbf{D}\bm{\xi} + \bm{\xi}\mathbf{D},
\end{equation}
and with Eq.~(\ref{eq:SimplificationAdjointRepresentation}), we apply the adjoint representation to both sides of Eq.~(\ref{eq:NewtonRhf}) to obtain
\begin{equation}
  [\mathbf{D},\mathbf{H}_{E}(\mathrm{ad}_{\mathbf{D}} (\mathbf{\Omega}))] 
    - [\mathbf{F}^{\mathrm{vv}}+\mathbf{F}^{\mathrm{oo}}, \mathrm{ad}_{\mathbf{D}} (\mathbf{\Omega})] 
    = -[\mathbf{D},\mathbf{F}],
  \label{eq:NewtonRhf2}
\end{equation}
where we introduced the occupied-occupied and virtual-virtual blocks of the AO basis representation of the Fock operator
\begin{align}
  \mathbf{F}^\mathrm{oo} &= \mathbf{D}\mathbf{F}\mathbf{D} \, ,
  \label{eq:FockOO} \\
  \mathbf{F}^\mathrm{vv} &= (\mathbbm{1}-\mathbf{D})\mathbf{F}(\mathbbm{1}-\mathbf{D}) \, .
  \label{eq:FockVV}
\end{align}
Note that, upon rearrangement, Eq.~(\ref{eq:NewtonRhf2}) is equivalent to the Newton equation given in Ref.~\citenum{Host2008_ARH}.
This enables us to reduce the dimension of the equations from $n^2$ to $m(n-m)$.
It is known that the eigenvalues of the density matrix are the occupation numbers which are either 0 or 1, which holds for any $\mathbf{D}\in\Gr$.
Therefore, there exists a unitary matrix, $\bm{\Theta}$, such that
\begin{equation}
    \mathbf{D} = \bm{\Theta} 
    \begin{pmatrix}
        \mathbbm{1}_m &   \bm{0}  \\
             \bm{0}        &   \bm{0}  \\
    \end{pmatrix}
    \bm{\Theta}^\mathrm{T}.
    \label{eq:diagonalDensityMatrix}
\end{equation}
The matrix $\bm{\Theta}$ can be chosen to be $\mathbf{C}$, the MO coefficient matrix, but different choices are possible.
In practice, any matrix $\bm{\Theta}=\mathbf{C}\mathbf{U}$, where $\mathbf{U}$ is a unitary matrix, still fulfills Eq.~(\ref{eq:diagonalDensityMatrix}) and $\mathbf{U}$ can be purposely chosen to enhance the algorithm convergence.
Eq.~(\ref{eq:diagonalDensityMatrix}) can be exploited to parametrize the adjoint representation at $\mathbf{D}$.
It can be verified that for any $\bm{\Omega}\in\mathfrak{so}_n$ and any $\mathbf{S}\in\mathrm{Sym}_n$, the following relations holds
\begin{align}
  \mathrm{ad}_{\mathbf{D}} (\mathbf{\Omega}) &=  \bm{\Theta} 
  \begin{pmatrix}
           \bm{0}               & \bm{\kappa}  \\
    \bm{\kappa}^\mathrm{T} &       \bm{0}      \\
  \end{pmatrix}
  \bm{\Theta}^\mathrm{T},
\\
  \mathrm{ad}_{\mathbf{D}} \left(\mathbf{S}\right) &= \bm{\Theta} 
  \begin{pmatrix}
              \bm{0}             & \bm{\kappa}  \\
    -\bm{\kappa}^\mathrm{T} &       \bm{0}      \\
  \end{pmatrix}
  \bm{\Theta}^\mathrm{T} \, ,
    \label{eq:BlockStructure} 
\end{align}
with $\bm{\kappa}\in \mathbb{R}^{m\times (n-m)}$.
By applying the Jacobi identity, the second term of Eq.~(\ref{eq:NewtonRhf2}) can be expressed as
\begin{equation}
  \begin{aligned}
    \mathrm{ad}_{\mathbf{D}} \big([\mathbf{F}^{\mathrm{vv}}+\mathbf{F}^{\mathrm{oo}}, \mathbf{\Omega}]\big)
    &=[\mathbf{F}^{\mathrm{vv}}+\mathbf{F}^{\mathrm{oo}}, \mathrm{ad}_{\mathbf{D}} (\mathbf{\Omega})] 
    + [\mathrm{ad}_{\mathbf{D}} (\mathbf{F}^{\mathrm{vv}}+\mathbf{F}^{\mathrm{oo}}), \mathbf{\Omega}]\\ 
    &= [\mathbf{F}^{\mathrm{vv}}+\mathbf{F}^{\mathrm{oo}}, \mathrm{ad}_{\mathbf{D}} (\mathbf{\Omega})],
  \end{aligned}
  \label{eq:NewtonAfterJacobi}
\end{equation}
where the last term in the first line evaluates to zero.
Since all terms in the Newton equations, Eq.~(\ref{eq:NewtonRhf2}), are of the form $\mathrm{ad}_{\mathbf{D}} \left(\mathbf{S}\right)$, we can apply $\bm{\Theta}$ on both sides which yields a block structure in the form of Eq.~(\ref{eq:BlockStructure}).
Consequently, we only need to consider either the upper-right or lower-left block of the transformed Newton equations.
To obtain the form that only contains the upper-right block, we transform Eq.~(\ref{eq:NewtonRhf2}) by left-multiplying with the transpose of the first $m$ columns of $\bm{\Theta}$, denoted by $\bm{\Theta}_\mathrm{o}$, and by right-multiplication with the $(n-m)$-last columns of $\bm{\Theta}$, denoted by $\bm{\Theta}_\mathrm{v}$.
This yields a more compact form of the Newton equation, where the left- and right-hand side are elements of $\mathbb{R}^{m\times (n-m)}$, according to
\begin{equation}
   \bm{\Theta}_\mathrm{o}^\mathrm{T}\Big([\mathbf{D},\mathbf{H}_{E}(\mathrm{ad}_{\mathbf{D}} (\mathbf{\Omega}))]  - [\mathbf{F}^{\mathrm{vv}}+\mathbf{F}^{\mathrm{oo}}, \mathrm{ad}_{\mathbf{D}} (\mathbf{\Omega})]\Big)\bm{\Theta}_\mathrm{v} = -\bm{\Theta}_\mathrm{o}^\mathrm{T}[\mathbf{D},\mathbf{F}]\bm{\Theta}_\mathrm{v}.
   \label{eq:NewtonRhfTheta}
\end{equation}
This enables optimizing directly the $\bm{\kappa}\in \mathbb{R}^{m\times (n-m)}$ parameters as in Eq.~(\ref{eq:BlockStructure}).
Hence, we can carry out the optimization with the minimal number of free parameters.

We now derive a computationally more efficient update-scheme for the density matrix compared to the matrix exponential of Eq.~(\ref{eq:ExponentialMap}).
It was shown in Ref.~\citenum{helmke2007_NewtonGrassmann} that the matrix exponential can be replaced with the Q-factor of the QR-decomposition of $\mathbf{X}_\mathrm{AO}=\mathbbm{1}-\mathrm{ad}^2_{\mathbf{D}} (\bm{\Omega})$.
$\mathbf{X}_\mathrm{AO}$ displays the following block structure when expressed in the $\bm{\Theta}$-basis:
\begin{equation}
  \mathbf{X}_{\Theta} = \bm{\Theta}^\mathrm{T} ( \mathbbm{1}-\mathrm{ad}^2_{\mathbf{D}_i} (\bm{\Omega}))\bm{\Theta} =
  \begin{pmatrix}
    \mathbbm{1}_m& \bm{\kappa}  \\
    -\bm{\kappa}^\mathrm{T} & \mathbbm{1}_{n-m} \\
  \end{pmatrix} \, .
  \label{eq:XMatrixThetaBasis}
\end{equation}
This structure enables us to find explicit equations for the Q and R factors.
Because the Q-factor is unitary, for $\mathbf{X}_\mathrm{\Theta}=\mathbf{Q}\mathbf{R}$, it holds that
\begin{equation}
  \mathbf{X}_{\Theta}^\mathrm{T}\mathbf{X}_{\Theta}= \mathbf{R}^\mathrm{T}\mathbf{R} =
  \begin{pmatrix}
    \mathbbm{1}_m + \bm{\kappa}\bm{\kappa}^\mathrm{T} &                         \bm{0}                        \\
                \bm{0}                                & \mathbbm{1}_{n-m} + \bm{\kappa}^\mathrm{T}\bm{\kappa} \\
  \end{pmatrix} \, ,
  \label{eq:QROfXMatrix}
\end{equation}
and therefore, the diagonal blocks of the R-factor can be obtained for a given $\bm{\kappa}$ by Cholesky decomposition according to
\begin{align}
  \mathbf{R}^\mathrm{T}_\mathrm{o}\mathbf{R}_\mathrm{o} &= \mathbbm{1}_m + \bm{\kappa}\bm{\kappa}^\mathrm{T},\\
  \mathbf{R}^\mathrm{T}_\mathrm{v}\mathbf{R}_\mathrm{v} &=\mathbbm{1}_{n-m} + \bm{\kappa}^\mathrm{T}\bm{\kappa}.
\end{align}
Hence, the Q-factor is obtained from Eq.~(\ref{eq:XMatrixThetaBasis}) as
\begin{equation}
  \mathbf{Q} =
  \begin{pmatrix}
    \mathbf{R}^{-1}_\mathrm{o}& \bm{\kappa}\mathbf{R}^{-1}_\mathrm{v}  \\
      -\bm{\kappa}^\mathrm{T}\mathbf{R}^{-1}_\mathrm{o} & \mathbf{R}^{-1}_\mathrm{v} \\
  \end{pmatrix} \, .
  \label{eq:QMatrixExpression}
\end{equation}
We note here that the Q-factor always exists and, consequently, the Cholesky decomposition and matrix inversion can be performed without loss in numerical accuracy (the proof can be found in Ref.~\citenum{helmke2007_NewtonGrassmann}).
Finally, we transform the Q-factor back to the atomic orbital basis as
\begin{equation}
  \mathbf{Q}_{\mathrm{AO}} = \bm{\Theta}\ \mathbf{Q}\ \bm{\Theta}^\mathrm{T} \, ,
  \label{eq:Qfactor}
\end{equation}
such that we can update the density matrix as
\begin{equation}
  \mathbf{D}_{i+1} = \mathbf{Q}_\mathrm{AO}\ \mathbf{D}_i\ \mathbf{Q}_\mathrm{AO}^\mathrm{T}.
  \label{eq:QfactprMap}
\end{equation}
Lastly, we note that the entire derivation up to this point did not depend on the explicit form of the energy function, but only on the fact that it is a function of a matrix that has to satisfy the constraints of the Grassmann manifold.

\section{Generalization to unrestricted and nuclear-electronic wave functions}
\label{sec:theory2}

We proceed by generalizing the methods that we presented above to the unrestricted and nuclear-electronic methods.
In both cases, the energy function depends on multiple density matrices.
We first give a short introduction to the nuclear-electronic Hartree--Fock method for notational purposes.
Then, we derive the corresponding Newton equations and present the analytic expressions for the Fock matrix and the Hessian of the energy. 
We conclude this section by introducing the ARH method for approximating the energy Hessian and, consequently, reduce the computational scaling.

\subsection{Nuclear-electronic Hartree--Fock Theory}

The nuclear-electronic Hartree--Fock wave function, $\Psi_\mathrm{NE}$, depends explicitly on the coordinates of electrons and nuclei and is a product of appropriately symmetrized products of orbitals, $\Phi_I$, for each particle type $I$.\cite{nakai2007nuclear,Hammes-Schiffer2002,Feldmann2022_QuantumProton}
Those symmetrized products are Slater determinants and permanents for fermions and Bosons, respectively.
The corresponding wave function ansatz reads
\begin{equation}
  \Psi_\mathrm{NE} (\mathbf{r}_{1},\dots,\mathbf{r}_{N_\text{t}}) = \prod_I^{N_{\text{t}}} \Phi_{I} (\mathbf{r}_{I}) ~.
  \label{eq:ansatz}
\end{equation}
Here, $N_\text{t}$ denotes the number of particle types and $\mathbf{r}_I$ is the vector that contains the positions of all particles of type $I$. 
$\Phi_{I}$ is constructed as a properly symmetrized product of nuclear or electronic spin-orbitals, $\phi_{Isi}$,
\begin{equation}
  \Phi_{I}(\mathbf{r}_I) = \frac{1}{\sqrt{N_{I}!}} \mathcal{S}_\pm \left( \prod^{S_I}_s \prod_i^{N_{Is}} \phi_{Iis}(\mathbf{r}_{Iis}) \right),
  \label{eq:SingleTypeAnsatz}
\end{equation}
where $S_I$ is the spin quantum number of type $I$ and $\mathbf{r}_{Iis}$ is the coordinate of particle $i$ of type $I$ with spin-projection $s$.
$N_I$ is the number of particles of type $I$ and $N_{Is}$ is the number of particles of type $I$ with a given spin-projection $s$.
The (anti)symmetrization operator, $\mathcal{S}_\pm$, enforces the proper permutational symmetry of the product of orbitals.
In this work, we will focus on spin-$\frac{1}{2}$ fermions, but the methods can easily be extended to particle types with different spins.
For a non-relativistic nuclear-electronic Hamiltonian (see, e.g., Ref.~\citenum{Feldmann2022_QuantumProton}) the Hartree--Fock energy can be expressed as
\begin{equation}
 \begin{aligned}
    E_\mathrm{NE}(\{\mathbf{D}_{Is}\}) = \ &\sum^{N_\mathrm{t}}_I \sum_s^{S_I} \sum_{\mu\nu}^{N_{\mathrm{AO},I}} h_{I\mu\nu} D_{Is\nu\mu}
                                        + \frac{1}{2} \sum^{N_\mathrm{t}}_I \sum_s^{S_I} \sum_{\mu\nu\rho\sigma}^{N_{\mathrm{AO},I}} D_{Is\nu\mu} \Bar{V}_{I\mu\nu,I\rho\sigma} D_{Is\sigma\rho} \\
                                      & + \sum^{N_\mathrm{t}}_I \sum_{s<t}^{S_I} \sum_{\mu\nu\rho\sigma}^{N_{\mathrm{AO},I}}
                                          D_{Is\nu\mu} V_{I\mu\nu,I\rho\sigma} D_{It\sigma\rho} \\
                                      & + \sum^{N_\mathrm{t}}_{I<J} \sum_{s}^{S_I} \sum_{t}^{S_J}\sum_{\mu\nu}^{N_{\mathrm{AO},I}} \sum_{\rho\sigma}^{N_{\mathrm{AO},J}}
                                          D_{Is\nu\mu} V_{I\mu\nu,J\rho\sigma} D_{Jt\sigma\rho}.
 \end{aligned}
 \label{eq:NuclearElectronicEnergy}
\end{equation}
The energy optimization of the nuclear-electronic wave function reads
\begin{equation}
  \mathrm{min} 
  \Big\{ 
    E_\mathrm{NE}(\{\mathbf{D}_{Is}\})\ \Big|\
    \mathbf{D}_{Is} \in \mathbb{R}^{N_{\mathrm{AO},I}\times N_{\mathrm{AO},I}},\ 
    \mathbf{D}_{Is}^2=\mathbf{D}_{Is}=\mathbf{D}_{Is}^\mathrm{T} ,\ \mathrm{Tr}\left(\mathbf{D}_{Is}\right) = N_{Is},\ \forall\ I,s
  \Big\} \, .
  \label{eq:EnergyMinimizationNe}
\end{equation}
Hence, the density matrices of each particle type and spin, $\mathbf{D}_{Is}$, are elements of a Grassmann manifold $\Gr$ with $m=N_{Is}$ and $n=N_{\mathrm{AO},I}$.

\subsection{Optimization on product manifolds}

In the following section, we derive the Newton equations for Hartree--Fock wave functions that depend on an arbitrary number $r$ of density matrices.
In this way, the resulting equations can be applied to unrestricted BO or nuclear-electronic problems.
We define the product manifold, $\mathrm{Gr}^{\times}_{\mathbf{m},\mathbf{n}}$, as\cite{Curtef2012_ProductGrassmann}
\begin{equation}
  \mathrm{Gr}^{\times}_{\mathbf{m},\mathbf{n}} 
    = \left\{ (\mathbf{D}_{1},\dots,\mathbf{D}_r)\ |\ \mathbf{D}_{i}\ \in\  \mathrm{Gr}_{m_i,n_i},\ i=1,\dots,r \right\} \, ,
  \label{eq:ProductGrassmann}
\end{equation}
with
\begin{equation}
  (\mathbf{m},\mathbf{n}) = \big( (m_1,n_1),\dots,(m_r,n_r)  \big).
  \label{eq:ProductIndices}
\end{equation}
We note that also $\mathrm{Gr}^{\times}_{\mathbf{m},\mathbf{n}}$ is a manifold, and therefore, the Newton equations can be derived as above.
The unrestricted or nuclear-electronic energies of Eq.~(\ref{eq:UhfElectronicEnergy}) or Eq.~(\ref{eq:NuclearElectronicEnergy}), respectively, are defined as functions on the Cartesian product space of symmetric matrices 
\begin{equation}
  E: \mathrm{Sym}_\mathbf{n}=\mathrm{Sym}_{n_1}\times\dots\times\mathrm{Sym}_{n_r} \rightarrow \mathbb{R}.
  \label{eq:FunctionOfSymmetricMatrices}
\end{equation}
We define $\mathcal{E}$, the restriction of the energy function $E$ to the manifold, $\mathrm{Gr}^{\times}_{\mathbf{m},\mathbf{n}}$, as
\begin{equation}
  \mathcal{E}: \mathrm{Gr}^{\times}_{\mathbf{m},\mathbf{n}}  \rightarrow \mathbb{R}.
  \label{eq:FunctionOfProductManifold}
\end{equation}
Consequently, the gradient of $E$ is obtained as 
\begin{equation}
  \mathrm{grad}_{E}((\mathbf{D}_{1},\dots,\mathbf{D}_r)) = (\mathbf{F}_1,\dots,\mathbf{F}_r),
  \label{eq:Grad_E_NE}
\end{equation}
which can be interpreted as a vector containing the Fock matrices of each type and spin, evaluated at the point $\mathbf{D}=(\mathbf{D}_{1},\dots,\mathbf{D}_r)\in  \mathrm{Gr}^{\times}_{\mathbf{m},\mathbf{n}}$. 
The gradient of $\mathcal{E}$ is given as the projection of Eq.~(\ref{eq:Grad_E_NE}) onto the tangent space $T_\mathbf{D}\mathrm{Gr}^{\times}_{\mathbf{m},\mathbf{n}}$.
This projection can be performed component-wise such that $\mathbf{F}_i$ is projected onto $T_{\mathbf{D}_i}\mathrm{Gr}_{m_i,n_i}$, resulting in

\begin{equation}
  \mathrm{grad}_{\mathcal{E}}(\mathbf{D}) = \big(\mathrm{ad}^2_{\mathbf{D}_1}(\mathbf{F}_1),\dots,\mathrm{ad}^2_{\mathbf{D}_r}(\mathbf{F}_r)\big).
  \label{eq:GradMulticomponent}
\end{equation}
Similarly, the elements of the Hessian matrix of $E$ read
\begin{equation}
  E^{i\mu\nu,j\sigma\rho} = \frac{\partial^2 E}{\partial D_{i\mu\nu} \partial D_{j\sigma\rho}} = \left( \frac{\partial \mathbf{F}_j}{\partial D_{i\mu\nu}} \right)_{\sigma\rho},
  \label{eq:SecondDerivativeEgeneralCase}
\end{equation}
and the action of the Hessian as a linear operator is defined as
\begin{equation}
    \begin{aligned}
     \mathbf{H}_{E}(\mathbf{S}) : \mathrm{Sym}_\mathbf{n} &\rightarrow\mathrm{Sym}_\mathbf{n},\\
     \mathbf{S}=(\mathbf{S}_1,\dots,\mathbf{S}_r) &\mapsto \mathbf{H}_{E}(\mathbf{D})(\mathbf{S}) = \big(\mathbf{H}_{E_1}(\mathbf{D})(\mathbf{S}),\dots, \mathbf{H}_{E_r}(\mathbf{D})(\mathbf{S})\big) \, .
    \end{aligned}
\end{equation}
The elements $ \mathbf{H}_{E,i}$ are given as
\begin{equation}
  \mathbf{H}_{E,i}(\mathbf{D})(\mathbf{S}) = \sum_j^r \mathbf{G}_{i}(\mathbf{S}_j) \, ,
  \label{eq:HessianProduct}
\end{equation}
where we introduced
\begin{equation}
  \mathbf{G}_{i}(\mathbf{S}_j) = \sum_{\mu\nu}^{n_i} \frac{\partial \mathbf{F}_i}{\partial D_{j\mu\nu}} S_{j\mu\nu},
  \label{eq:GMatrix}
\end{equation}
which we will refer to as the $G$-matrix.
In the restricted case, the $G$-matrix coincides with the Hessian of $E$.
The Hessian of the energy on the manifold, $\mathcal{E}$, is obtained as
\begin{equation}
  \begin{aligned}
    \mathbf{H}_{\mathcal{E}}(\bm{\xi}) : T_\mathbf{D}\mathrm{Gr}^{\times}_{\mathbf{m},\mathbf{n}} &\rightarrow T_\mathbf{D}\mathrm{Gr}^{\times}_{\mathbf{m},\mathbf{n}},\\
    \bm{\xi}=(\bm{\xi}_1,\dots,\bm{\xi}_r) &\mapsto \mathbf{H}_{\mathcal{E}}(\mathbf{D})(\bm{\xi}) = \big(\mathbf{H}_{\mathcal{E}_1}(\mathbf{D})(\bm{\xi}),\dots, \mathbf{H}_{\mathcal{E}_r}(\mathbf{D})(\bm{\xi})\big),
  \end{aligned}
  \label{eq:MulticomponentHessian}
\end{equation}
with
\begin{equation}
  \mathbf{H}_{\mathcal{E},i}(\mathbf{D})(\bm{\xi}) = \mathrm{ad}^2_{\mathbf{D}_i} \big(\mathbf{H}_{E,i}(\mathbf{D})(\bm{\xi})  \big) 
                                                   - \mathrm{ad}_{\mathbf{D}_i} \mathrm{ad}_{\mathbf{F}_i  } (\bm{\xi}_i).
\end{equation}
This enables us to write the unrestricted/nuclear-electronic Newton equations as
\begin{equation}
  \mathrm{ad}^2_{\mathbf{D}_i} \big(\mathbf{H}_{E,i}(\mathbf{D})(\bm{\xi})  \big) - \mathrm{ad}_{\mathbf{D}_i} \mathrm{ad}_{\mathbf{F}_i} (\bm{\xi}_i)
    =  -\mathrm{ad}^2_{\mathbf{D}_i}(\mathbf{F}_i) \quad \mathrm{for}\ i=1,\dots,r.
  \label{eq:NewtonGeneral}
\end{equation}
Eq.~(\ref{eq:NewtonGeneral}) represents $r$ equations that are coupled via $\bm{\xi}_j\in T_{\mathbf{D}_j}\mathrm{G}_{m_j,n_j}$ for $j=1,\dots,r$.
By applying $\mathrm{ad}_{\mathbf{D}_i}$ on both sides of Eq.~(\ref{eq:NewtonGeneral}), we obtain
\begin{equation}
  \sum_j^r[\mathbf{D}_i,\mathbf{G}_{i}(\mathrm{ad}_{\mathbf{D}_j} (\bm{\xi}_j))] 
     - [(\mathbf{F}_i^{\mathrm{vv}}+\mathbf{F}_i^{\mathrm{oo}}), \mathrm{ad}_{\mathbf{D}_i} (\bm{\xi}_i)]
    = -[\mathbf{D}_i,\mathbf{F}_i].
  \label{eq:NewtonGeneral2}
\end{equation}
As above, we can apply a unitary transformation $\bm{\Theta}_i$ to the equations in such a way that they can be recast into the previously introduced block structure.
The density matrices can then be updated as described above.

\subsection{Exact Newton nuclear-electronic Hartree--Fock optimization}

The Newton equations, Eq.~(\ref{eq:NewtonRhf}) and (\ref{eq:NewtonGeneral}), do not depend on the specific form of the energy function.
We now calculate the Fock matrix and the Hessian for our specific target, i.e., the Hartree--Fock energy function.
To keep the derivation as general as possible, we will focus on the unrestricted nuclear-electronic case.
The Fock matrices are obtained as the derivatives of Eq.~(\ref{eq:NuclearElectronicEnergy}) as
\begin{equation}
 \begin{aligned}
  F_{Is\mu\nu} = \frac{\partial E}{\partial D_{Is\nu\mu}} =\ 
    &h_{I\mu\nu} + \sum_{\rho\sigma}^{N_{\mathrm{AO},I}} \Bar{V}_{I\mu\nu,I\rho\sigma} D_{Is\sigma\rho}
                 + \sum_{t\neq s}^{S_I} \sum_{\rho\sigma}^{N_{\mathrm{AO},I}} V_{I\mu\nu,I\rho\sigma} D_{It\sigma\rho} \\
    &+ \sum^{N_\mathrm{t}}_{J\neq I} \sum_{t}^{M_{S_J}} \sum_{\rho\sigma}^{N_{\mathrm{AO},J}} V_{I\mu\nu,J\rho\sigma} D_{Jt\sigma\rho} \, .
 \end{aligned}
 \label{eq:FockMatrix}
\end{equation}
Next, we rewrite the Fock matrices in a compact form in terms of the same particle type-same spin coupling matrix, $B_{Is\mu\nu}$,
\begin{equation}
  B_{Is\mu\nu} = \sum_{\rho\sigma}^{N_{\mathrm{AO},I}} \Bar{V}_{I\mu\nu,I\rho\sigma} D_{Is\sigma\rho},
  \label{eq:SpinCouplingMatrix}
\end{equation}
the matrix for the same particle type-different spin coupling, $J_{It\mu\nu}$,
\begin{equation}
  J_{It\mu\nu} = \sum_{\rho\sigma}^{N_{\mathrm{AO},I}} V_{I\mu\nu,I\rho\sigma} D_{It\sigma\rho},
  \label{eq:DifferentSpinCoupling}
\end{equation}
and the different particle type coupling matrix, $L_{I,Jt\mu\nu}$,
\begin{equation}
  L_{I,Jt\mu\nu} = \sum_{\rho\sigma}^{N_{\mathrm{AO},J}}
  V_{I\mu\nu,J\rho\sigma} D_{Jt\sigma\rho}.
  \label{eq:NEcouplingMatrix}
\end{equation}
With these definitions, the Fock matrix element can be written compactly as
\begin{equation}
 \begin{aligned}
  F_{Is\mu\nu} = h_{I\mu\nu}
    + B_{Is\mu\nu}
    + \sum_{t\neq s}^{S_I} J_{It\mu\nu}
    + \sum^{N_\mathrm{t}}_{J\neq I} \sum_{t}^{S_J} L_{I,Jt\mu\nu}.
 \end{aligned}
 \label{eq:NuclearElectronicFockMatrix}
\end{equation}
To calculate the Hessian, we must calculate the second derivative of the energy function with respect to a density matrix element of the same type and spin, the same type but different spin, and a different particle type.
This yields the following matrix elements:
\begin{align}
    \frac{\partial F_{Is\mu\nu}}{\partial D_{Is\sigma\rho}} &= \Bar{V}_{I\mu\nu,I\rho\sigma} ,\\
    \frac{\partial F_{Is\mu\nu}}{\partial D_{It\sigma\rho}} &= V_{I\mu\nu,I\rho\sigma} ,\\
    \frac{\partial F_{Is\mu\nu}}{\partial D_{Jt\sigma\rho}} &= V_{I\mu\nu,J\rho\sigma} .
\end{align}
Finally, the $G$-matrices can be expressed in terms of the quantities defined above as
\begin{equation}
  \begin{aligned}
  G_{Is\mu\nu}(\bm{\xi}_{Is}) &=\sum_{\rho\sigma}^{N_{\mathrm{AO},I}} \Bar{V}_{I\mu\nu,I\rho\sigma} \xi_{Is\sigma\rho},\\
  G_{Is\mu\nu}(\bm{\xi}_{It}) &= \sum_{\rho\sigma}^{N_{\mathrm{AO},I}} V_{I\mu\nu,I\rho\sigma} \xi_{It\sigma\rho},\\
  G_{Is\mu\nu}(\bm{\xi}_{Jt}) &= \sum_{\rho\sigma}^{N_{\mathrm{AO},J}}
  V_{I\mu\nu,J\rho\sigma} \xi_{Jt\sigma\rho} \, .
 \end{aligned}
 \label{eq:NEcoupling}
\end{equation}
Note that Eq.~(\ref{eq:NEcoupling}) closely resembles the $\mathbf{B}_{Is}$, $\mathbf{J}_{It}$, and $\mathbf{L}_{I,Jt}$ matrices from Eqs.~(\ref{eq:SpinCouplingMatrix})--(\ref{eq:NEcouplingMatrix}).
The $G$-matrices can, therefore, be evaluated with the same computer routines that build the Fock matrices.
Evaluating the $G$- and Fock matrices is the main bottleneck associated with the solution of Newton equations that scales as $\mathcal{O}(N_\mathrm{AO}^4)$.
The $G$- and Fock matrices enter the exact nuclear-electronic Hessian according to
\begin{equation}
  \begin{aligned}
     \mathbf{H}_{\mathcal{E},Is}(\mathbf{D})(\bm{\xi})
       =&\  -[ \mathbf{F}^{\mathrm{oo}}_{Is} + \mathbf{F}^{\mathrm{vv}}_{Is}, \bm{\xi}_{Is}]
       +  \sum_{J} \sum_{t} [ \mathbf{D}_{Is},\mathbf{G}_{Is}(\bm{\xi}_{Jt})] \, .
  \end{aligned}
  \label{eq:ExactNeHessian}
\end{equation}
The Newton optimization of the Hartree--Fock energy is performed iteratively in two layers which are the macro- and microiterations.
During the microiterationss, the tangent vector $\bm{\xi}$ is optimized by solving the Newton equations.
The density matrix is then updated, which concludes a single macroiteration.
The new set of Newton equations is then solved.
While the Fock matrix must be evaluated only once per macroiteration, the $G$-matrices explicitly depend on the tangent vectors and, therefore, they must be recalculated at every microiteration.

\subsection{The Augmented Roothaan--Hall method}
\label{subsec:ARH}

The ARH method has been introduced by H{\o}st et al.\ \cite{Host2008_ARH} to approximate the Hessian, Eq.~(\ref{eq:ExactNeHessian}), such that its evaluation cost in every microiteration is effectively reduced to $\mathcal{O}(N_\mathrm{AO}^3)$.
To this end, we first define the relative direction of the density matrix at the current, $n$-th iteration, $\mathbf{D}^{n}_{Is}$, with respect to the one at the $i$-th previous iteration, $\mathbf{D}^{i}_{Is}$, as:
\begin{equation}
  \mathbf{D}^{in}_{Is} = \mathbf{D}^{i}_{Is} - \mathbf{D}^{n}_{Is}.
  \label{eq:densityMatrixDifference}
\end{equation}
The $G$-matrices can be exactly expressed as
\begin{align}
  \mathbf{G}_{Is}(\mathbf{D}^{in}_{Is}) &= \mathbf{B}^{i}_{Is}-\mathbf{B}^{n}_{Is}= \mathbf{B}^{in}_{Is},
  \label{eq:KMatrix} \\
  \mathbf{G}_{Is}(\mathbf{D}^{in}_{It}) &= \mathbf{J}^{i}_{It}-\mathbf{J}^{n}_{It} =\mathbf{J}^{in}_{It},\\
  \mathbf{G}_{Is}(\mathbf{D}^{in}_{Jt}) &= \mathbf{L}^{i}_{I,Jt}-\mathbf{L}^{n}_{I,Jt} =\mathbf{L}^{in}_{I,Jt}.
  \label{eq:ApproxNEcoupling} 
\end{align}
Based on these exact expressions of the $G$-matrices, the Hessian at the given iteration can be approximated in terms of quantities calculated in the previous iterations.
Following H{\o}st et al., \cite{Host2008_ARH} we introduce the basis, $\mathcal{B}_{Is,n}$, spanned by the \textit{directions}, Eq.~(\ref{eq:densityMatrixDifference}), to the previous iterations,
\begin{equation}
  \mathcal{B}_{Is,n} = \{\ket{i_{Is}}= \mathbf{D}^{in}_{Is}\},
  \label{eq:BasisARH}
\end{equation}
with the corresponding metric 
\begin{equation}
  T_{Is,ij} = \braket{i_{Is}|j_{Is}}= \Tr\left(\mathbf{D}^{in}_{Is}\mathbf{D}^{jn}_{Is}\right) \, .
  \label{eq:MetricTensor}
\end{equation}
The key approximation underlying the ARH method is to assume that $\mathrm{Span}(\mathcal{B}_{Is,n}) \approx \mathrm{Sym}_{N_{AO,I}}$.
Since $\mathrm{Sym}_{N_{AO,I}}$ is spanned by $N_{AO,I}(N_{AO,I}+1)/2$ independent parameters, this assumption will be exact only after the same number of iterations, provided that the iterations generate linearly independent density matrices.
However, a reliable approximation to the Hessian is obtained with a significantly smaller subspace.
If the approximation holds, we can expand an arbitrary matrix $\mathbf{S}_{Is}=\ket{s_{Is}}\in\mathrm{Sym}_{N_{AO,I}}$ in this basis as
\begin{equation}
  \ket{s_{Is}} = \mathbbm{1}\ket{s_{Is}} = \sum_{ij} \ket{i_{Is}} (\mathbf{T}^{-1}_{Is})_{ij} \braket{j_{Is}|s_{Is}}.
  \label{eq:ARH_expansion_of_vector}
\end{equation}
For the purpose of the derivation, we explicitly introduce the basis of independent pairs of AOs as $\mathcal{B}_{Is,\mathrm{AO}}=\{\ket{\mu}\}$ which exactly spans $\mathrm{Sym}_{N_{AO,I}}$.
The density matrices, the Fock matrices, and the integrals are expressed in this basis.
The directions $\ket{i_{Is}}$ can be expanded in this basis as
\begin{equation}
  \ket{i_{Is}} = \sum_{\mu} \ket{\mu_{Is}} \braket{\mu_{Is}|i_{Is}} = \sum_{\mu} D^{in}_{Is\mu} \ket{\mu_{Is}} \, ,
  \label{eq:DirectionInAOBasis}
\end{equation}
and the $G$-matrices (here in the same-type same-spin case) can be expressed as operators acting on this space
\begin{equation}
  \widehat{G}_{Is,Is} = \sum_{\mu\nu} \ket{\mu_{Is}} {E}^{Is\mu,Is\nu} \bra{\nu_{Is}} \, ,
  \label{eq:GMatrixSubspace}
\end{equation}
where $ {E}^{Is\mu,Is\nu}$ is defined as in Eq.~(\ref{eq:SecondDerivativeEgeneralCase}).
Now we can express the action of $\widehat{G}_{Is,Is}$ on the matrix $\ket{s_{Is}}$ according to
\begin{equation}
 \begin{aligned}
    \widehat{G}_{Is,Is}\ket{s_{Is}} 
     &= \sum_{\mu\nu}  \ket{\mu_{Is}} {E}^{Is\mu,Is\nu} \braket{\nu_{Is}|\mathbbm{1}|s_{Is}} \\
     &= \sum_{\mu\nu}  \sum_{ij}  \ket{\mu_{Is}} {E}^{Is\mu,Is\nu} 
        D^{in}_{Is\nu} (\mathbf{T}^{-1}_{Is})_{ij} \braket{j_{Is}|s_{Is}} \\
     &= \sum_{\mu}     \sum_{ij}  B^{in}_{Is\mu} (\mathbf{T}^{-1}_{Is})_{ij} 
        \braket{j_{Is}|s_{Is}} \ket{\mu_{Is}} \, ,
 \end{aligned}
\end{equation}
where we have exploited Eqs.~(\ref{eq:KMatrix}) and (\ref{eq:ARH_expansion_of_vector}).
We can then express the $G$-matrices at the $n$-th iteration entirely in terms of the quantities introduced above as
\begin{align}
  \mathbf{G}_{Is}^{n}(\mathbbm{1}\bm{\xi}_{Is}) &= \sum_{ij} \mathbf{B}^{in}_{Is}   (\mathbf{T}^{-1}_{Is})_{ij} \Tr(\mathbf{D}^{jn}_{Is}\bm{\xi}_{Is}) \, , \\
  \mathbf{G}_{Is}^{n}(\mathbbm{1}\bm{\xi}_{It}) &= \sum_{ij} \mathbf{J}^{in}_{It}   (\mathbf{T}^{-1}_{Is})_{ij} \Tr(\mathbf{D}^{jn}_{Is}\bm{\xi}_{It}) \, , \\
  \mathbf{G}_{Is}^{n}(\mathbbm{1}\bm{\xi}_{Jt}) &= \sum_{ij} \mathbf{L}^{in}_{I,Jt} (\mathbf{T}^{-1}_{Is})_{ij} \Tr(\mathbf{D}^{jn}_{Is}\bm{\xi}_{Jt}) \, ,
\end{align}
where the evaluation scales as $\mathcal{O}(N_\mathrm{AO}^3)$.
With the approximate $G$-matrices, we write the ARH Hessian at the $n$-th iteration as 
\begin{equation}
  \begin{aligned}
     \mathbf{H}^\mathrm{ARH}_{\mathcal{E},Is}(\mathbf{D}^n)(\bm{\xi}^n)
         =&  -[ \mathbf{F}^{\mathrm{oo},n}_{Is} + \mathbf{F}^{\mathrm{vv},n}_{Is}, \bm{\xi}^n_{Is}]
             + \sum_{ij}  [ \mathbf{D}^n_{Is},\mathbf{B}^{in}_{Is}] (\mathbf{T}^{-1}_{Is})_{ij}
                          \Tr(\mathbf{D}^{jn}_{Is}  \bm{\xi}^n_{Is} ) \\
           & + \sum_{t\neq s}\sum_{ij}  [ \mathbf{D}^n_{Is},\mathbf{J}^{in}_{It}] (\mathbf{T}^{-1}_{Is})_{ij}      
                                          \Tr(\mathbf{D}^{jn}_{Is} \bm{\xi}^n_{It} ) \\
           & + \sum_{J\neq I} \sum_{t} \sum_{ij}  [ \mathbf{D}^n_{Is},\mathbf{L}^{in}_{I,Jt}] (\mathbf{T}^{-1}_{Is})_{ij}    
                                                    \Tr(\mathbf{D}^{jn}_{Is}  \bm{\xi}^n_{Jt}) \, .
  \end{aligned}
  \label{eq:ARHEquation}
\end{equation}

\section{Nuclear-electronic trust-region augmented Hessian method}
\label{sec:trah}

Iteratively solving the exact or approximate Newton equations leads to a convergence to the closest stable point of any kind, i.e., minimum, maximum, or saddle point.
Convergence to a minimum is, however, only guaranteed when the Hessian is positive definite. 
This condition will be satisfied if the region in which the truncated Taylor expansion is accurate contains a minimum. 
This is often not the case in the first iteration steps where one is usually far away from the minimum.
We overcome this issue with the trust-region method.\cite{Absil2009_OptOnMatrixManifolds,Fletcher2013_PracticalOptimization}
The general idea is to locate a region around the current approximate solution in which the quadratic model is a good approximation and to find a positive-definite symmetric matrix that is sufficiently close to the exact Hessian.
This idea is implemented by introducing the trust radius, which is the maximum length of the optimization step that is still contained in the trust region.
The Hessian is then level-shifted such that the resulting matrix is positive-definite.

In order to constrain the Newton equations within the trust region, we first define the vector $\bm{\kappa}$ that contains the orbital update matrices of all types and spins
\begin{equation}
  \bm{\kappa} = 
  \begin{pmatrix}
    \mathrm{vec}(\bm{\kappa}_{1})   \\
    \mathrm{vec}(\bm{\kappa}_{2}) \\
    \vdots
 \end{pmatrix},
\end{equation}
where $\mathrm{vec}(\mathbf{M})$ is a vector that collects the entries of a matrix $\mathbf{M}$.
The vector form of the gradient of Eq.~(\ref{eq:GradMulticomponent}) in the ${\Theta}$-basis is defined as
\begin{equation}
  \mathbf{g} =
    \begin{pmatrix}
      \mathrm{vec}\big(\bm{\Theta}^\mathrm{T}_{1,\mathrm{o}}
      \mathrm{ad}_{\mathbf{D}_1}(\mathbf{F}_1)\bm{\Theta}_{1,\mathrm{v}}\big) \\
      \mathrm{vec}\big(\bm{\Theta}^\mathrm{T}_{2,\mathrm{o}}
      \mathrm{ad}_{\mathbf{D}_2}(\mathbf{F}_2)\bm{\Theta}_{2,\mathrm{v}}\big) \\
      \vdots
    \end{pmatrix}.
  \label{eq:gradientfin}
\end{equation}
We write the Hessian of type $i$, introduced in Eq.~(\ref{eq:MulticomponentHessian}), as a function of $\bm{\kappa}$, as
\begin{equation}
  \mathbf{H}_i(\bm{\kappa})=\mathbf{H}_{\mathcal{E}_i}(\mathbf{D})(\bm{\xi}),
  \label{eq:HessianKappa}
\end{equation}
where the parameters $\bm{\xi}_i$ are given as
\begin{equation}
  \bm{\xi}_i = \bm{\Theta}_i
  \begin{pmatrix}
    0& \bm{\kappa}_i  \\
    \bm{\kappa}_i^\mathrm{T} & 0 \\
  \end{pmatrix}
  \bm{\Theta}_i^\mathrm{T}.
  \label{eq:XiParametrization}
\end{equation}
Hence, we can write the vector form of the Hessian of the entire system in the $\bm{\Theta}$-basis as
\begin{equation}
  \mathbf{H}(\bm{\kappa}) =
    \begin{pmatrix}
      \mathrm{vec}\big(\bm{\Theta}^\mathrm{T}_{1,\mathrm{o}}
      \mathbf{H}_1(\bm{\kappa}) \bm{\Theta}_{1,\mathrm{v}}\big)
      \\
      \mathrm{vec}\big(\bm{\Theta}^\mathrm{T}_{2,\mathrm{o}}
      \mathbf{H}_1(\bm{\kappa}) \bm{\Theta}_{2,\mathrm{v}}\big)\\
      \vdots
    \end{pmatrix}.
  \label{eq:HessianFin}
\end{equation}
Now, we can rewrite the Newton equations, Eq.~(\ref{eq:NewtonGeneral}), in a compact form as
\begin{equation}
  \mathbf{H}(\bm{\kappa}) = -\mathbf{g}.
  \label{eq:NewtonSimple}
\end{equation}
Our next aim is twofold: we want (i) to enforce that the solution vector $\bm{\kappa}$ lies within the trust region and that (ii) the orbital-update step is directed towards the next minimum.
We can achieve this \textit{via} the augmented Hessian approach where we introduce the free parameters $\alpha$, $\mu$, and the augmented Newton equations 
\begin{equation}
  \mathbf{H}(\alpha)
    \begin{pmatrix}
      1\\
      \mathbf{k}(\alpha)
    \end{pmatrix}
    =
    \mu
    \begin{pmatrix}
      1\\
      \mathbf{k}(\alpha)
    \end{pmatrix},
  \label{eq:trahEquations}
\end{equation}
where the scaled augmented Hessian reads
\begin{equation}
\mathbf{H}(\alpha)=
  \begin{pmatrix}
    0 & \alpha \mathbf{g}^T \\
    \alpha\mathbf{g} & \mathbf{H}
  \end{pmatrix}.
  \label{eq:ScaledAH}
\end{equation}
Eq.~(\ref{eq:trahEquations}) is equivalent to the following set of equations:
\begin{align}
  \alpha \mathbf{g}^\mathrm{T} \mathbf{k} &= \mu,\\
  (\mathbf{H}-\mu\mathbbm{1}) \mathbf{k}(\alpha) &= -\alpha\mathbf{g},
\end{align}
where
\begin{equation}
  \bm{\kappa}(\mu) =\alpha^{-1} \mathbf{k}(\alpha)
\end{equation}
solves the level-shifted Newton equations
\begin{equation}
  (\mathbf{H}-\mu\mathbbm{1}) \bm{\kappa}(\mu) = -\mathbf{g}.
\end{equation}
It can be shown\cite{Hoyvik2012_TrahOrbitalLocalization} that the lowest eigenvalue of the augmented Hessian is lower than the lowest eigenvalue of the Hessian. Therefore, if we select $\mu$ as the lowest eigenvalue of the augmented Hessian, $(\mathbf{H}-\mu\mathbbm{1})$ will be positive-definite and the optimization will be directed towards a minimum.
The $\alpha$-parameter is adjusted such that the step-length constraint is fulfilled, i.e.,
\begin{equation}
    \lVert 
    \bm{\kappa}(\mu)\rVert_2 \leq h,
    \label{eq:trustRadiusConstraint}
\end{equation}
where $\lVert\cdot\rVert_2$ denotes the $\ell^2$-norm and $h$ is the trust radius.
This step-length constraint can be implemented, e.g., with the bisection method as described in Appendix~\ref{sec:AppendixTRAH}.
For a detailed account of the augmented Hessian method see, e.g., Ref.~\citenum{Hoyvik2012_TrahOrbitalLocalization}.
We give the details of our implementation in Appendix~\ref{sec:AppendixTRAH}.

\section{Computational Details}
\label{sec:details}

All calculations presented in the next section rely on the def2-SVP and def2-TZVP Karlsruhe\cite{Ahlrichs2005_def2} electron basis sets.
For the protons, we employ the (solid-harmonic) protonic basis (PB) sets, PB4-D = \{4s3p2d\} and PB4-F1 = \{4s3p2d1f\} by Hammes-Schiffer and coworkers.\cite{Hammes-Schiffer2020_NuclearBasis}
We denote the combined electronic (X) and protonic (Y) basis sets as [$\el$:(X),$\pr$:(Y)].
We implemented the theory presented in this work in our new \texttt{C++} software library, named \texttt{Kiwi} which is integrated in our open-source \texttt{SCINE} framework\cite{scine}.
The \texttt{Kiwi} library relies on \texttt{Libint}\cite{Libint2} for the integral evaluation.
The convergence criterion that we adopt in the orbital optimization is that the gradient norm falls below $10^{-5}$.

\subsection{Born--Oppenheimer calculations}
\label{subsec:BO_details}

We first apply the algorithms presented above to unrestricted electronic Hartree--Fock calculations. 
Our test cases will be the iron-sulfur clusters [Fe$_2$S$_2$(SCH$_3$)$_4$]$^{2-}$ and [Fe$_4$S$_4$(SCH$_3$)$_4$]$^{2-}$, abbreviated in the following as [2Fe--2S] and [4Fe--4S], respectively, with the geometries taken from Ref.~\citenum{Sharma2014_IronSulfur}.
We also report results obtained with the diagonalization-based RH optimization as implemented in Orca 5.0.2., with the PModel starting guess and the TightSCF settings.\cite{Neese2020_Orca} 
By default, Orca accelerates the SCF convergence with a combination of static damping,\cite{Hartree1947_ScfDamping} level-shifting,\cite{Saunders1973_ScfLevelShift} and DIIS.\cite{Pulay1980_DIIS} 
We activate the SlowConv or VerySlowConv keywords when the optimization does not converge with the default settings.
The Newton and ARH optimizations are carried out with our \texttt{Kiwi} program.
We construct the initial guess as superposition of atomic densities\cite{Almlof1982_SAD} (SAD) with the algorithm described in Ref.~\citenum{Kansik2009_ARH3L_SAD}.
The SAD guess produces a density matrix that is not an element of the Grassmann manifold.
This impedes to straightforwardly apply the second-order methods.
Therefore, we construct a Fock matrix from the SAD density matrix and run a single RH iteration to ensure that the initial density matrix is valid. 
The simulation settings are provided in Appendix~\ref{sec:AppendixTRAH}. 
Only for the Newton optimization, we reduced the number of Davidson iterations to 16 to avoid unnecessary micro-iterations.
Moreover, we perturbed the initial Davidson guess with random noise to break its restricted symmetry and steer the optimization towards an unrestricted solution, as suggested in Ref.~\citenum{Helmich2021_OrcaTRAH}.
For all calculations, we characterize the final stationary point by calculating the lowest eigenvalue of the Hessian, Eq.~(\ref{eq:HessianFin}).
A negative eigenvalue indicates convergence to a saddle point, while a minimum will be reached if the lowest eigenvalue is positive.

\subsection{Nuclear-electronic calculations}
\label{subsec:NE_details}

We apply the nuclear-electronic variants of the algorithms introduced above to the water clusters H$_5$O$_2^+$, H$_9$O$_4^+$, H$_{11}$O$_5^+$, and H$_{13}$O$_6^+$, with geometries taken from Ref.~\citenum{Kulig2013_WaterClusters}, and to H$_3$O$^+$ with the structure taken from Ref.~\citenum{Kulig2014_WaterClusters2}.
We demote the oxygen nuclei as point charges and treat the protons quantum mechanically.
We approximate the electronic wave function with restricted Hartree--Fock and assume the high-spin approximation for the nuclei, i.e., we set the total nuclear spin to the maximal positive value. We note here that we need to scale the restricted block of the Hessian by a factor of two in order to guarantee that the Hessian is symmetric when mixing restricted and unrestricted particle types.

We carry out nuclear-electronic ARH and Newton calculations with \texttt{Kiwi} and the simulation details are given in Appendix~\ref{sec:AppendixTRAH}.
We compare the results with these obtained with the DIIS algorithm\cite{Pulay1980_DIIS} and with the geometric direct minimization (GDM) \cite{Voorhis2002geometric,Dunietz2002_GDM} method as implemented in \texttt{Kiwi} and Q-Chem 5.4 \cite{QChem5_2021}, respectively.
In the latter case, we applied the default convergence settings.
With these criteria, the converged energies obtained with the two programs agree at least up to $0.1$~$\upmu\text{Ha}$ for the same molecule.

Both the GDM and the diagonalization-based RH implementations in Q-Chem and in our \texttt{Kiwi} program apply the alternating optimization strategy: starting with the protons, the RH equations of a specific particle type are optimized in the mean-field potential generated by the constant density matrices of all other particle types.
Once convergence or the maximum number of nested iterations is reached, the RH equations associated with the next particle type are solved.
We construct the initial electronic guess for the nuclear-electronic Hartree--Fock optimization from converged orbitals obtained from a BO Hartree--Fock calculation.
To construct the protonic guess density matrix, we have employed two different procedures. The first is the nuclear-electronic analog to the core-matrix guess. This guess is obtained by diagonalizing the protonic core matrix, $\mathbf{h}_{\pr}$, that comprises the kinetic energy and the Coulomb interaction with the classical nuclei (i.e., the one-body contributions to the Fock operator).
Alternatively, we construct the nuclear guess from a superposition of nuclear densities (SND) obtained from an atomic nuclear-electronic calculation.\cite{Almlof1982_SAD,VanLenthe2006_SND}
To that end, we construct the molecular nuclear density matrix as a direct sum of the atomic density matrices obtained from an atomic SCF optimization.
Note that, in the atomic calculations, we subtract the center-of-mass contribution from the Hamiltonian as in Ref.~\citenum{Muolo2020_Nuclear}.
We observed that the core-matrix guess leads to a consistently slower convergence rate.
Therefore, if not stated otherwise, we apply the SND guess.

\section{Results and Discussion}
\label{sec:results}

\subsection{Iron-sulfur clusters}
\label{subsec:IronSolfur}

In this section, we report results of the unrestricted Hartree--Fock energy optimization of the iron-sulfur clusters [2Fe--2S] and [4Fe--4S].
The electronic ground state of both clusters was theoretically determined to be a singlet.\cite{Sharma2014_IronSulfur}
However, multireference calculations usually rely on orbitals calculated for the high-spin state, because the corresponding Hartree--Fock optimization is much easier to converge.
For [2Fe--2S], the high spin state corresponds to a multiplicity of 11 ($S_z=5$) and for [4Fe--4S], the multiplicity is 21 ($S_z=10$). 
We report in Table~\ref{tab:FeS} the number of microiterations required to converge the Newton method while, for the other methods, we report the number (macro)iterations.
In this way, we count the number of iterations with a computational step that scales as $\mathcal{O}(N_\mathrm{AO}^4)$ for every method. 
In the Newton method the Fock matrix is constructed once per macroiteration, while the exact Hessian is evaluated at every microiteration.
Those steps are the bottlenecks of the optimization algorithm.
As for the ARH and RH methods, the Fock matrix evaluation is instead the limiting step.

\begin{table}
  \caption{Number of iterations required to converge the unrestricted Hartree--Fock energy for iron-sulfur clusters in the low-spin and high-spin state with the def2-SVP basis set.
  We report the number of microiterations for the Newton method and the (macro)iterations for all other methods.\\
  The [2Fe--2S] system in this basis posses 572 free parameters, and the [4Fe--4S] system 768.\\
  Calculations marked as $\ddagger$ converged to a saddle point.\\
  $^*$ Carried out with Orca 5.0.2.\cite{Neese2020_Orca}.}
  \label{tab:FeS}
  \centering
  \begin{tabular}{ l l l l l}
    \hline\hline
    Method      &  [2Fe--2S]&& [4Fe--4S]& \\
                & Iterations & Energy/Ha & Iterations & Energy/Ha            \\
    \hline
    low-spin &  & & &\\
    \hline
    RH$^*$    & 2205$^\ddagger$   & $-5067.050 207$  & 1990$^\ddagger$   & $-8386.268 907$  \\
    ARH                             & 125               & $-5067.300 886$  & 167               & $-8386.604 242$  \\
    Newton & 386               & $-5067.300 886$  & 446               & $-8386.616 887$  \\
    \hline
    high-spin & & & & \\
    \hline
    RH$^*$    & 18                & $-5067.504 791$  & 641               & $-8386.855 264$  \\
    ARH                             & 21                & $-5067.504 793$  & 56                & $-8386.854 139$  \\
    Newton & 75                & $-5067.504 792$  & 251               & $-8386.854 139$  \\
    \hline\hline
  \end{tabular}
\end{table}

Table~\ref{tab:FeS} highlights that converging the low-spin state is particularly hard for both clusters.
The diagonalization-based RH method converges only after 2000 iterations with static damping, level-shifting, and DIIS. 
Moreover, a stability analysis reveals that, for both clusters, the RH optimization converged to a saddle point.
The ARH method reduces the computational cost for both clusters by more than an order of magnitude compared to the RH method. 
Additionally, both solutions are local minima, as confirmed by the stability analysis.
The Newton method takes approximately three times more iterations than ARH but remains more efficient than the RH method.
For the [4Fe--4S] cluster, the Newton method converges to a lower-energy minimum than ARH.
We note here that none of the algorithms can guarantee to find the global minimum but, instead, they will converge to the local minimum that is the closest to the initial guess.
Depending on the complexity of the parameter landscape, the optimization may therefore be very sensitive to the initial guess. 
Moreover, if multiple local minima are close to the initial guess, ARH and the Newton method may converge to different stationary points --- which is the effect we observe here.
To show this, we have perturbed the initial orbitals for the ARH optimization with the orbital steering method\cite{Vaucher2017_OrbitalSteering}.
The orbital steering randomly selects pairs of orbitals and rotates them with a randomly selected small angle.
This perturbs the initial guess by breaking the symmetry between spin-up and spin-down orbitals, leading to a final energy of $-8386.670390~\text{Ha}$ that is converged after 167 iterations.
Hence, a small random perturbation of the initial guess is sufficient to converge to an energy value that is lower by 50~mHa.
This substantiates our initial assumption that the energy landscape has a complex structure with multiple local minima close to the initial guess.

Table~\ref{tab:FeS} shows that the orbital optimization is easier to converge for the high-spin case than for the low-spin one.
In this case, the RH method converges for the [2Fe-2S] cluster to a local minimum in the least number of iterations followed by ARH and, then, the Newton method.
For the [4Fe--4S] cluster, the RH method takes the largest number of iterations, while the ARH method outperforms it by at least an order of magnitude.
Note that the Newton method converges in a smaller number of iterations compared to the RH method. 
In this case, all algorithms converge to local minima and the RH method finds the lowest energy solution.
As we already noted above, this discrepancy is the result of the sensitivity of the optimization algorithm to the initial guess.

We conclude this section by reporting the averaged timing of the first 50 optimization steps for [2Fe--2S] with the the ARH method and the RH method with DIIS as implemented in our in-house code for comparability.
We carried out the calculations on a single core with an Intel(R) Xeon(R) Gold 6136 CPU.
The averaged time of a single ARH iteration is 21.25 seconds and for RH with DIIS it is 20.52 seconds.
Especially for large systems, when the evaluation of the Fock matrix and its diagonalization become the bottleneck, we expect that the performance of our ARH method will surpass the one of the RH method. 

To summarize, in all cases but one, the RH method yields the slowest convergence rate and it converged to saddle points in two calculations. 
The ARH method, with approximately the same computational cost per iteration as the RH method, takes the least number of iterations and was in three cases more efficient than the RH method by at least one order of magnitude.
Moreover, ARH always converged to local minima.
The Newton method is more efficient than the RH method in the two cases that are the most difficult to optimize.
However, it is consistently slower than ARH and, therefore, does not provide an advantage over the latter.
Note that we carried out all ARH and Newton optimizations with the very same parameter set reported in Appendix~\ref{sec:AppendixTRAH}.
Conversely, for the RH method, we tested the Orca settings for normal, slow, and very slow convergence until a solution was found. 
This highlights the robustness of both the ARH and the Newton methods.

\subsection{Nuclear-electronic description of water clusters}

\begin{table}[htbp!]
  \caption{Number of iterations required to optimize the nuclear-electronic Hartree--Fock energy based on the def2-TZVP electronic basis set and the PB4-D protonic one.
  We report the number of microiterations for the Newton method and the macroiterations for all other methods. \\
  $^*$ Carried out with Q-Chem 5.4.2.\cite{QChem5_2021}. }
  \label{tab:PreBO_Convergence}
  \centering
  \begin{tabular}{ l c c c c c}
    \hline\hline
    Method      &  H$_3$O$^+$ & H$_5$O$_2^+$ & H$_9$O$_4^+$ & H$_{11}$O$_5^+$ & H$_{13}$O$_6^+$  \\
    \hline
    Number of orbitals      & 118 & 207  & 385  &   474   &      563 \\
    \hline
    GDM$^*$                         &     450     &     533      &     875      &       833       &      1100       \\
    DIIS                            &     161     &     173      &     187      &       195       &      194        \\
    Newton &     120     &     146      &     172      &       174       &      171        \\
    ARH                             &     47      &     49       &     68       &       69        &      64         \\
    \hline
    Energy/Ha    &     $-76.225709$     &     $-152.258071$       &     $-304.274 597$       &       $-380.271584$        &       $-456.273 674 $        \\
    \hline\hline
  \end{tabular}
\end{table}

We now apply the DIIS, GDM, ARH, and Newton orbital optimization algorithms to water clusters of increasing size.
As we show in Table~\ref{tab:PreBO_Convergence}, the efficiency of these algorithms is drastically different already for these relatively small molecules.
We employed the [$\el$:(def2-TZVP),$\pr$:(PB4-D)] basis set and treated all protons quantum mechanically.
As in the previous case, we report the number of (macro)iterations for the RH, GDM, and ARH methods, and the number of microiterations for the Newton method.
Table~\ref{tab:PreBO_Convergence} indicates that the convergence rate depends heavily on the algorithm.
The GDM method yields consistently the slowest convergence, while the ARH method outperforms all other methods.
It improves upon the GDM method by more than an order of magnitude and it requires approximately 3 to 3.5 times less iterations to converge than DIIS.
Moreover, while the number of iterations of the GDM method increases with the system size, the convergence rate of the other methods mildly depends on it.
Our results indicate that the ARH method is the most efficient strategy for optimizing the nuclear-electronic energy.
It converges significantly faster than GDM and DIIS, while exploiting Hessian information and with a computational cost that is smaller than that of the full Newton optimization.
Moreover, we observe that the nuclear-electronic Hessian can have some remaining small negative eigenvalues at convergence.
However, note that every optimization method converges to the same energy independent of the guess (see Table~\ref{tab:PreBO_Convergence}), and the magnitude of these negative eigenvalues is -- in absolute values -- consistently small.
These two facts indicate that the optimization landscape is very shallow and, therefore, the lowering in energy along the directions corresponding to the negative eigenvalues is negligible.

\begin{figure}[htbp!]
  \centering
  \includegraphics[width=0.7\textwidth]{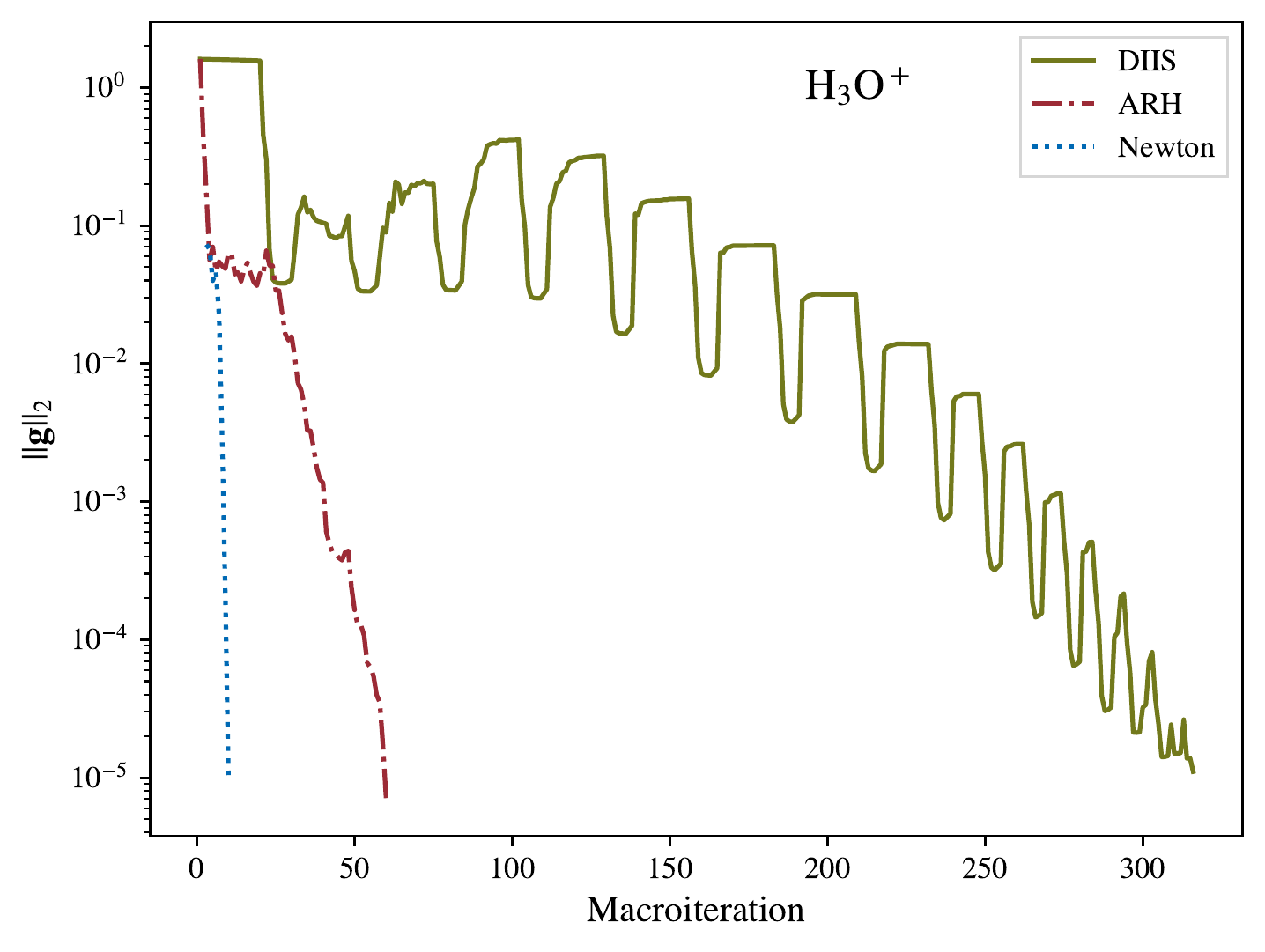}
  \caption{Convergence of the gradient $\ell^2$-norm with the number of macroiterations with the DIIS, ARH, and exact Newton method for H$_3$O$^+$ with the [$\el$:(def2-TZVP),$\pr$:(PB4-F1)] basis set.
  The calculations started from the core-matrix guess for the protons.}
  \label{fig:h3o_conv}
\end{figure}

Next, with the example of H$_3$O$^+$ and the [$\el$:(def2-TZVP),$\pr$:(PB4-F1)] basis set, we report in Figure~\ref{fig:h3o_conv} the convergence of the $\ell^2$-norm of the gradient with respect to the number of iterations with the Newton, ARH, and DIIS methods.
Note that from here on, we also report the number of macroiterations for the Newton method.
Moreover, we chose the core-matrix guess as a starting point for the protonic density matrix.
With the Newton method, convergence was reached after 10 macroiterations, while the ARH method converged after 60 iterations.
The DIIS method converges after 316 iterations.
During the first few iterations, the curves of the Newton and ARH method are almost identical.
Because the ARH subspace is too small to reliably approximate the Hessian at the beginning of the optimization, this suggests that its contribution is negligible.
After the first few iterations, the ARH curve deviates from the Newton one and takes more macroiterations to converge.
This indicates that the energy landscape becomes more complex and an accurate representation of the Hessian is required to achieve optimal convergence.
The DIIS curve strongly deviates from both the Newton and the ARH one, and shows significant oscillations at the first iterations which decrease only towards the end of the optimization.
The oscillations are a result of the alternating optimization strategy.
In fact, the orbitals of a given particle type --- say, the electrons --- are first optimized for a given set of orbitals of the protons.
The latter are then optimized by fixing the orbitals of the electrons.
This second step will inevitably lead to an increase of the norm of the electronic gradient.

\begin{figure}[htbp!]
  \centering
    \centering
    \includegraphics[width=0.7\textwidth]{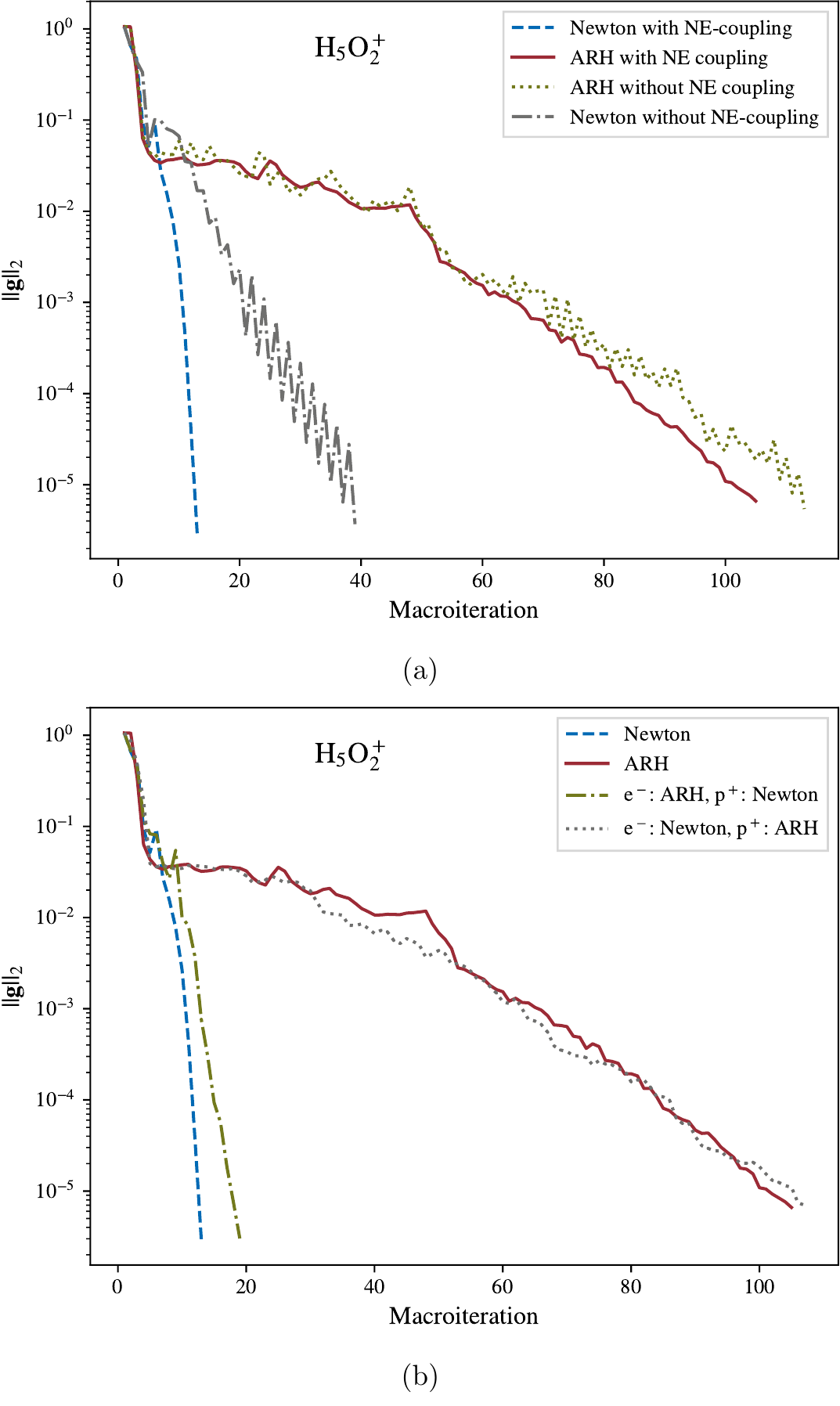}
  \caption{Convergence of the gradient $\ell^2$-norm (a) with and without including the nuclear-electronic (NE) coupling and (b) with exact and approximate Hessian contribution for H$_5$O$_2^+$ with the [$\el$:(def2-TZVP),$\pr$:(PB4-F1)] basis set.
  We note that the calculations started from a core-matrix guess for the protons.}
  \label{fig:ARH_vs_secondOrder}
\end{figure}

We investigate in Figure~\ref{fig:ARH_vs_secondOrder}a the influence of the nuclear-electronic coupling on the convergence of the Newton and ARH methods with the example of H$_5$O$_2^+$ with the [$\el$:(def2-TZVP),$\pr$:(PB4-F1)] basis set, starting from the core-matrix guess for the protons. 
Here, the nuclear-electronic coupling refers to the exact or approximate $G$-matrices from Eq.~(\ref{eq:NEcoupling}) or (\ref{eq:ApproxNEcoupling}), respectively.
When the coupling is neglected, the nuclear-electronic blocks of the Hessian are zero.
Hence, the exact or approximate nuclear and electronic Newton equations are not coupled anymore.
Figure~\ref{fig:ARH_vs_secondOrder}a shows the convergence of the overall gradient norm with the number of macroiterations.
The Newton method converges smoothly in less than 20 iterations when the coupling is included.
However, when this coupling is neglected, the number of iterations increases to 40, and the norm of the gradient oscillates significantly during the optimization.
Instead, neglecting the nuclear-electron coupling does not alter significantly the convergence rate of ARH.
The optimization without the nuclear-electronic coupling takes only 8 iterations more (113) compared to the one with the coupling (105) and only weak oscillations in the gradient norm are observed.

\begin{figure}[htbp!]
  \centering
  \includegraphics[width=0.7\textwidth]{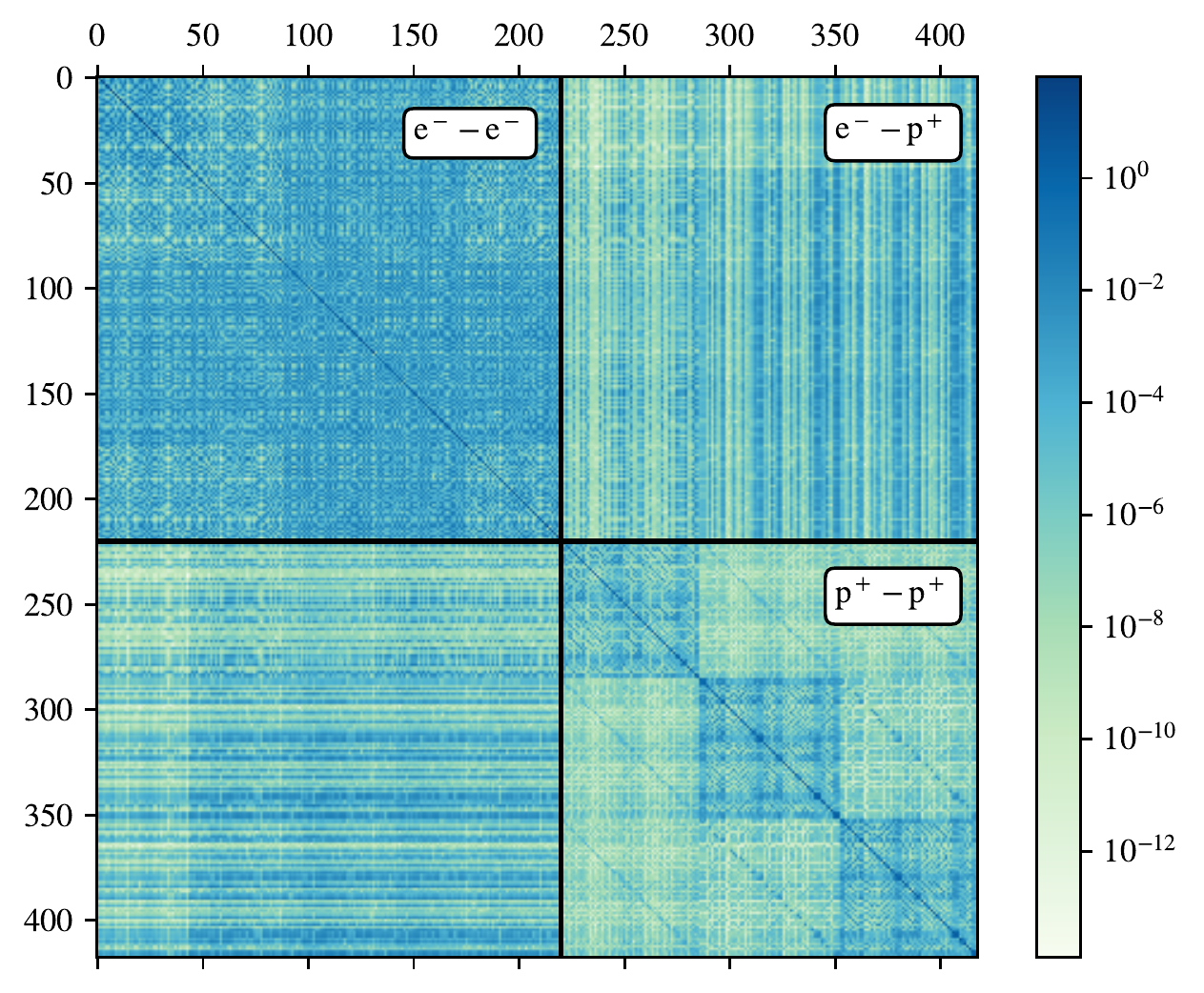}
  \caption{Absolute values of the virtual-occupied blocks of the exact Hessian at convergence of H$_3$O$^+$ with the [$\el$:(def2-TZVP),$\pr$:(PB4-D)] basis set.
  Shown is a logarithmic color map for the electron (upper left), proton (lower right), and the electron-proton (off-diagonal) blocks.}
  \label{fig:h3o_hessian}
\end{figure}

To further investigate the contribution of the nuclear-electronic coupling elements to the Hessian, we show in Figure~\ref{fig:h3o_hessian} the absolute values of the entries of the virtual-occupied blocks of the exact Hessian of H$_3$O$^+$ with the [$\el$:(def2-TZVP),$\pr$:(PB4-D)] basis set.
We report the graphical representation of the Hessian for the protonated water molecule because, for larger molecules, the Hessian matrix would be too large to be represented in a single figure.
The electron-electron block is dense, the proton-proton block a block-sparse structure, and the electron-proton block shows a band-structure.
Therefore, the coupling among the electronic orbitals is stronger than the proton-electron or proton-proton coupling.
The band structure in the electron-proton block shows that the electron-proton coupling is strong only for a few nuclear orbitals.
Moreover, the electron-proton coupling is in general not significantly smaller than electron-electron or proton-proton coupling.

To understand why the electron-proton coupling --- although of comparable magnitude to the electron-electron and proton-proton coupling --- does not play a decisive role in the ARH optimization, we refer to Figure~\ref{fig:ARH_vs_secondOrder}b.
The figure shows the influence of calculating the exact Hessian or the ARH one on the convergence of the gradient norm with the number of iterations for H$_5$O$_2^+$ with the [$\el$:(def2-TZVP),$\pr$:(PB4-F1)] basis set. 
We report the ARH and Newton curves, together with results obtained with the exact protonic Hessian and the approximate electronic ARH one, and vice versa.
This means that we apply Eq.~(\ref{eq:ARHEquation}) to the orbital rotations of one particle type and Eq.~(\ref{eq:ExactNeHessian}) to the other one.
Optimizing the electrons with the exact instead of the ARH equation has almost no influence on the optimization.
Conversely, by optimizing electrons with ARH and protons with the Newton method, the convergence is almost as fast as with the exact Newton method (13 vs. 19 iterations, respectively).
Hence, the ARH method accurately approximates the electronic energy landscape but not the protonic one.
This indicates that the protonic landscape has a higher degree of non-convexity than the electronic one, i.e., the proton-proton block of the Hessian is the hardest to approximate.
This also explains why neglecting the nuclear-electronic block of the Hessian has only a minor effect on the convergence of both the Newton and the ARH optimization schemes, as shown in Figure~\ref{fig:ARH_vs_secondOrder}a.
Therefore, the slow convergence of nuclear-electronic Hartree--Fock compared to the purely electronic one does not result from the inter-particle-type coupling, but from the complexity of the nuclear energy landscape.

\section{Conclusions}
\label{sec:conclusions}

In this work, we have developed a differential-geometry-based framework that facilitates the derivation of second-order nuclear-electronic orbital optimization algorithms.
This framework enabled us to extend and efficiently implement the exact Newton and ARH algorithms for unrestricted (nuclear-)electronic Hartree--Fock calculations.
Our implementation reduces heavily any computational overhead for ARH when compared to RH with DIIS. 
Moreover, the evaluation of the ARH equations can be implemented with the same efficient incremental building and integral-screening techniques as in RH.
These techniques cannot be leveraged for exact Newton optimization schemes, which require the construction of the exact energy Hessian.
We showed that ARH is more stable and efficient than RH-DIIS for both strongly-correlated electronic and nuclear-electronic systems --- in some cases even by an order of magnitude.
Our second-order orbital optimization strategies are of particular advantage for strongly-correlated systems, i.e., in the presence of near-degenerate molecular orbitals and a vanishing HOMO-LUMO gap.
In these cases, the Hartree--Fock energy landscape has a particularly high degree of non-convexity and, therefore, the SCF procedure is hard to converge with conventional optimization methods.
We further reduced the computational cost by introducing the SND guess.
This guess was developed in analogy to a guess known from electronic-structure theory, namely, the SAD guess. For the SND guess, we optimize the Hydrogen atom without the BO approximation and use a direct sum of replicas of the converged Hydrogen atom protonic density matrix as the initial guess density matrix. 

Our results indicate why the convergence rate of nuclear-electronic Hartree--Fock calculations is significantly slower than in the electronic-only analog.
In Ref.~\citenum{Valeev2004_ENMO}, which is the only work that addresses this issue, the reason for the poor convergence was claimed to be the non-separability of the orbital-based nuclear-electronic wave function into the center-of-mass and internal components.
This, however, turned out not to be a relevant factor for the calculations presented in this work, because we treated at least one nucleus as a point charge fixed in space.
This breaks the translational symmetry of the Hamiltonian and, therefore, the wave function separability.
Our results suggest that the parametrized energy landscape defined in terms of the nuclear wave function parameters is significantly more complex than for the electronic-only analog, which we consider the ultimate reason for the slow convergence of nuclear-electronic Hartree--Fock calculations.

We note that, when different SCF optimization algorithms yield different Hartree--Fock energies, they produce different molecular orbitals.
This will affect any post-SCF calculation that relies on these molecular orbitals.
Naturally, those orbitals associated with the lowest energy may be considered the most reliable starting point in general.
Post-Hartree--Fock methods that do not account for orbital relaxation effects, such as second--order M{\o}ller--Plesset perturbation theory and Coupled Cluster, will be particularly sensitive to the convergence of the SCF procedure.
By contrast, methods that re-optimize the orbitals will mitigate the effect of a molecular orbital basis not associated with the lowest energy.
The algorithms introduced in this paper enable calculating the Hessian and, therefore, discriminate whether the SCF optimization has converged to a minimum or to a saddle point.
In the latter case, the molecular orbitals can be perturbed along the directions defined by eigenvectors associated with negative eigenvalues.
The SCF optimization can then be restarted from the resulting guess to converge to orbitals representing an energy minimum in parameter space.
In order to further discriminate whether the resulting minimum is global or local, the orbital steering algorithm introduced in Ref.~\citenum{Vaucher2017_OrbitalSteering} can be employed.
The orbital steering method randomly selects pairs of molecular orbitals and rotates them by a random angle.
This enables exploring a wider portion of the energy landscape and, therefore, to avoid convergence into local minima.

In our previous work,\cite{Feldmann2022_QuantumProton} we showed that nuclear-electronic correlation may be strong also for systems with a weakly correlated BO electronic wave function.
Similar results were recently obtained by Brorsen\cite{Brorsen2020_SelectedCI-PreBO}.
In addition, we have shown\cite{Feldmann2022_QuantumProton} that the entanglement between the nuclear orbitals among themselves is stronger than that between the electronic orbitals among themselves.
Both the complexity of the nuclear energy landscape and these strong correlation effects may stem from the fact that Gaussian-type basis functions are poorly suited\cite{Cassam-Chenai2021_BasisFunctionENMF} to describe nuclear orbitals, which also explains why the convergence to the complete basis set limit is slower for the nuclei even though they are highly localized compared to the electrons.\cite{Feldmann2022_QuantumProton}
Finally, we note that, although we focused on nuclear-electronic Hartree--Fock theory, the framework presented in this work can be straightforwardly extended to nuclear-electronic density functional theory considered in Ref.~\citenum{Hammes-Schiffer2012_MulticomponentDFT}.

\section*{Acknowledgements}
R.~F. is grateful to the G\"unthard Foundation for a PhD scholarship.
M.~R. and A.~B. are grateful for generous support through the ``Quantum for Life Center'' funded by the Novo Nordisk Foundation (grant NNF20OC0059939).

\section*{Appendix}

\subsection*{Implementation of the trust-region augmented Hessian method}
\label{sec:AppendixTRAH}

Our implementation of the trust-region augmented Hessian method is inspired by Refs.~\citenum{Salek2007_LinearScalingHF}, \citenum{Hoyvik2012_TrahOrbitalLocalization}, and\ \citenum{Helmich2021_OrcaTRAH}, with some modifications.
Following Helmich-Paris,\cite{Helmich2021_OrcaTRAH} we solve the augmented Hessian equations, Eq.~(\ref{eq:trahEquations}), and the Newton equations, Eq.~(\ref{eq:NewtonSimple}), with the Davidson method. 
We implemented the Davidson algorithm as in Ref.~\citenum{Crouzeix1994_Davidson}, with the trust-region augmented Hessian-specific modifications, and extended it to solve linear systems of equations.

The Davidson method aims at obtaining the lowest (or highest) few eigenvalues and vectors of a matrix by projecting it into a subspace --- built from the so-called trial-vectors --- that is generated by the Rayleigh-Ritz procedure.
The subspace is iteratively enlarged until it approximates the target eigenvectors with the desired accuracy.\cite{Crouzeix1994_Davidson}
Moreover, it is crucial for the Davidson algorithm to be efficient that the matrix to be diagonalized is diagonally-dominant and that an effective pre-conditioner is found.
The first point is addressed in our work by choosing a suitable $\bm{\Theta}$, see Eq.~(\ref{eq:diagonalDensityMatrix}).

\subsection*{Basis for the Davidson subspace}

We introduce the $n$-dimensional trial-basis, $\mathbf{B}_n$, and the $(n+1)$-dimensional augmented trial-Basis, $\mathbf{B}_n^\mathrm{A}$, which span the Davidson subspace.
The bases are, in the minimal case, one- or two-dimensional, respectively, and are written as
\begin{equation}
  \mathbf{B}_n^\mathrm{A} = 
  \begin{pmatrix}
       1   &      0       &       0       & \dots\\
    \bm{0} & \mathbf{b}_1 &  \mathbf{b}_2 & \dots\\
  \end{pmatrix}
  =
  \begin{pmatrix}
      1    &   \bm{0}    \\
    \bm{0} & \mathbf{B}_n\\
  \end{pmatrix},
  \label{eq:DavidsonSubspace}
\end{equation}
which will ensure that the projected Hessian has a special structure that we will exploit as explained below.
We highlight that we construct explicitly only $\mathbf{B}_n$, but not $\mathbf{B}_n^\mathrm{A}$.
The first trial-vector, $\mathbf{b}_1$, is defined as
\begin{equation}
  \mathbf{b}_1 = \mathbf{g}\ \lVert \mathbf{g} \rVert_2^{-1},
  \label{eq:FirstTrialVector}
\end{equation}
with $\mathbf{g}$ defined in Eq.~(\ref{eq:gradientfin}).
We accelerate the convergence by adding a second and third initial vector defined as
\begin{equation}
  \mathbf{b}_2 = \mathbf{g} + \mathbf{H}(\mathbf{g}),
  \label{eq:SecondTrialVector}
\end{equation}
with $\mathbf{H}(\cdot)$ as defined in Eq.~(\ref{eq:HessianFin}).
$\mathbf{b}_3$ is the vector containing only zeros except for the element with the occupied-virtual index associated with the largest difference between an occupied and virtual diagonal element of the $\bm{\Theta}$-transformed Fock matrix, which is set to one.
The trial vectors are orthonormalized with respect to $\mathbf{b}_1$ with the modified Gram--Schmidt procedure.
This guarantees that the augmented Hessian, Eq.~(\ref{eq:trahEquations}), has the following structure in the trial vector basis
\begin{equation}
  \mathbf{A}^{B^A_n} = 
  \begin{pmatrix}
              0                       & \alpha\lVert \mathbf{g} \rVert_2 &      0        & \dots   &     0        \\
    \alpha\lVert \mathbf{g} \rVert_2  &              H^{B_n}_{11}        & H^{B_n}_{12}  & \dots   & H^{B_n}_{1n} \\
              0                       &              H^{B_n}_{21}        & H^{B_n}_{22}  & \dots   & H^{B_n}_{2n} \\
           \vdots                     &                 \vdots           &    \vdots     & \ddots  &   \vdots     \\
              0                       &              H^{B_n}_{n1}        & H^{B_n}_{n2}  & \dots   & H^{B_n}_{nn} \\
  \end{pmatrix}
  =
  \begin{pmatrix}
            0                  &  \alpha(\mathbf{g}^{B_n})^\mathrm{T} \\
    \alpha \mathbf{g}^{B_n}    &             \mathbf{H}^{B_n}         \\
  \end{pmatrix} \, ,
  \label{eq:AugmentedHessianBlockStructure}
\end{equation}
where we defined the subspace-gradient, $\mathbf{g}^{B_n}$, as
\begin{equation}
  \mathbf{g}^{B_n} = 
  \begin{pmatrix}
    \alpha\lVert \mathbf{g} \rVert_2 \\
                 0                   \\
               \vdots 
  \end{pmatrix} \, ,
  \label{eq:SubspaceGradient}
\end{equation}
and the projected Hessian, $\mathbf{H}^{B_n}$, as
\begin{equation}
  \mathbf{H}^{B_n} = 
  \begin{pmatrix}
    H^{B_n}_{11} & H^{B_n}_{12} & \dots    & H^{B_n}_{1n} \\
    H^{B_n}_{21} & H^{B_n}_{22} & \dots    & H^{B_n}_{2n} \\
         \vdots  &   \vdots     & \ddots   &   \vdots     \\
    H^{B_n}_{n1} & H^{B_n}_{n2} & \dots    & H^{B_n}_{nn} \\
  \end{pmatrix} \, .
\end{equation}
The entries of $\mathbf{H}^{B_n}$ are obtained as
\begin{equation}
  H^{B_n}_{ij} = \mathbf{b}_i^\mathrm{T} \mathbf{H}(\mathbf{b}_j) = \mathbf{b}_i^\mathrm{T} \bm{\sigma}_j,
  \label{eq:ARH_Entries}
\end{equation}
where we defined the sigma-vector $\bm{\sigma}_j$ as $ \mathbf{H}(\mathbf{b}_j)$.
We define $\bm{\Sigma}_n$ as the matrix collecting all sigma-vectors as columns to write the Hessian in the trial-vector basis compactly as
\begin{equation}
  \mathbf{H}^{B_n} = \mathbf{B}_n^\mathrm{T} \bm{\Sigma}_n.
  \label{eq:HessianTrialBasis}
\end{equation}

\subsection*{Davidson algorithm in the global region}

As we discussed in Section~\ref{sec:trah}, we apply the trust-region method in the global region of the optimization, where the quadratic approximation is not accurate and does not contain a minimum.
Therefore, we solve the eigenvalue problem in the reduced space according to
\begin{equation}
  \mathbf{A}^{B^A_n}
  \begin{pmatrix}
              1              \\
    \mathbf{k}^{B_n}(\alpha) \\
  \end{pmatrix} 
  = \mu
  \begin{pmatrix}
             1               \\
    \mathbf{k}^{B_n}(\alpha) \\
  \end{pmatrix} \, ,
  \label{eq:SubspaceLevelShiftedNewton}
\end{equation}
where the level-shifted Newton equations, Eq.~(\ref{eq:LevelShiftedNewton}), are solved by
\begin{equation}
  \bm{\kappa}^{B_n}(\mu) = \alpha^{-1}\mathbf{k}^{B_n}(\alpha) \, .
  \label{eq:LevelShiftedNewton}
\end{equation}
We solve the eigenvalue problem with the algorithm implemented in the Eigen library.\cite{eigenweb}
We highlight here that the ARH Hessian is not necessarily exactly symmetric and, consequently, no self-adjoint eigensolver can be employed.
If the step-length constraint is not fulfilled, we adjust $\alpha$ at each Davidson iteration step such that Eq.~(\ref{eq:trustRadiusConstraint}) is fulfilled.
Since the Davidson basis is orthonormal, the constraint is fulfilled if
\begin{equation}
  \lVert  \bm{\kappa}^{B_n}(\mu)\rVert_2 \leq h \, . 
  \label{eq:DavidsonConstraint}
\end{equation}
If Eq.~(\ref{eq:DavidsonConstraint}) does not apply for $\alpha=1$, an upper bound $\alpha_\mathrm{max}$ is determined such that the constraint is fulfilled.
In our implementation, we start with $\alpha_\mathrm{max}=10$ and then iteratively add increasing powers of $10$ to $\alpha_\mathrm{max}$.
Thereafter, $\alpha$ is adjusted with the bisection algorithm such that the step length is equal to the boundary of the trust radius, i.e., $\lVert\bm{\kappa}^{B_n}(\mu)\rVert_2 = h$.
If an appropriate $\alpha$ is found, the solution vector in the original basis, $\bm{\kappa}_n(\mu)$, at the current Davidson iteration is given as
\begin{equation}
  \bm{\kappa}_n(\mu) = \mathbf{B}_n \bm{\kappa}^{B_n}(\mu).
  \label{eq:DavidsonSolution}
\end{equation}
The accuracy of the solution vector $\bm{\kappa}_n(\mu)$ at the $n$-th iteration is calculated as the $\ell^2$-norm of the residual, $\mathbf{r}_n$, which is given as
\begin{equation}
  \mathbf{r}_n = \mathbf{g} + (\bm{\Sigma}_n \bm{\kappa}^{B_n}(\mu)- \mu \bm{\kappa}_n(\mu)) \, .
  \label{eq:ErrorVector}
\end{equation}
The iterative diagonalization is terminated when the error falls below the numerical threshold.
Otherwise, a new trial vector is added to the basis.
For a given preconditioning matrix, $\bm{\Lambda}$, the new trial vector is obtained as:

\begin{equation}
  \mathbf{b}_{n+1} = (\bm{\Lambda}-\mu\mathbbm{1})^{-1}\mathbf{r}_n \, ,
  \label{eq:SubspaceExpansion}
\end{equation}
which is then Gram--Schmidt orthonormalized to all previous trial-vectors and added to $\mathbf{B}_n\rightarrow \mathbf{B}_{n+1}$.
In the $\bm{\Theta}$-basis, the preconditioner $\bm{\Lambda}$ contains the differences between the virtual and occupied diagonal elements of the Fock matrix in the $\bm{\Theta}$-basis, which is a good approximation to the eigenvalues of the Hessian.
We summarize the overall algorithm in Algorithm~\ref{alg:davidson_global}.

\begin{algorithm}[htbp!]
  \caption{Davidson algorithm in the global region.}
  \label{alg:davidson_global}
  \begin{algorithmic}
    \State \textbf{Compute} {$\mathbf{B}_{n_\mathrm{ini}}$, $\bm{\Sigma}_{n_\mathrm{ini}}$, $\mathbf{A}^{B^A_{n_\mathrm{ini}}}$, $\bm{\Lambda}$}
    \For{n:=$n_\mathrm{ini}$,$n_\mathrm{ini}+1$, $\dots$ }
      \State $\alpha \gets 1$
      \State \textbf{Solve:} $\mathbf{k}^{B_n}(\alpha=1) \gets$ Eq.~(\ref{eq:SubspaceLevelShiftedNewton})
      \If{$\lVert \alpha^{-1} \mathbf{k}^{B_n}(\alpha)\rVert_2 > h$} 
        \State \textbf{Find:}  $\mathbf{k}^{B_n}(\alpha_\mathrm{max}),\alpha_\mathrm{max}$ for which $\lVert \alpha_\mathrm{max}^{-1} \mathbf{k}^{B_n}(\alpha_\mathrm{max})\rVert_2 < h$
        \State \textbf{Find:}  $\mathbf{k}^{B_n}(\alpha_\mathrm{opt}),\alpha_\mathrm{opt} \in [1,\alpha_\mathrm{max}]$ such that $\lVert \alpha_\mathrm{opt}^{-1} \mathbf{k}^{B_n}(\alpha_\mathrm{opt})\rVert_2 = h$
        \State $\alpha \gets \alpha_\mathrm{opt}$
      \EndIf
      \State $\bm{\kappa}^{B_n}(\mu) \gets \alpha^{-1}\mathbf{k}^{B_n}(\alpha)$
      \State $\bm{\kappa}_n(\mu) \gets  \mathbf{B}_n \bm{\kappa}^{B_n}(\mu)$
      \State $ \mathbf{r}_n \gets \mathbf{g} + (\bm{\Sigma}_n \bm{\kappa}^{B_n}(\mu)- \mu \bm{\kappa}_n(\mu))$
      \If{$\lVert\mathbf{r}_n\rVert_2\leq \mathrm{threshold}$}
        \State exit for loop.
      \EndIf
      \State $\mathbf{b}_{n+1} \gets (\bm{\Lambda}-\mu\mathbbm{1}) ^{-1}\mathbf{r}_n$
      \State \textbf{Update:} $\mathbf{B}_{n+1}$, $\bm{\Sigma}_{n +1}$, $\mathbf{A}^{B^A_{n+1}}$
    \EndFor
  \end{algorithmic}
\end{algorithm}

\subsection*{Davidson algorithm in the local region}

In the local region, we assume that the quadratic model of the energy is valid and, therefore, we directly solve the Newton equations, Eq.~(\ref{eq:NewtonSimple}).
Those equations are often solved with the preconditioned conjugate-gradient method.\cite{Larsen2001_AODensityMatrixOptimization,Salek2007_LinearScalingHF,Host2008_ARH}
However, we found that a Davidson-like approach is more stable and faster, as also noted by Helmich-Paris.\cite{Helmich2021_OrcaTRAH}
In the Davidson subspace, the Newton equations read
\begin{equation}
  \mathbf{H}^{B_n} \bm{\kappa}^{B_n}=-\mathbf{g}^{B_n},
  \label{eq:SubspaceNewton}
\end{equation}
which we solve with the LU decomposition with full pivoting as implemented in Eigen \cite{eigenweb}.
The residual is obtained as
\begin{equation}
  \mathbf{r}_n = \mathbf{g} + \bm{\Sigma}_n\bm{\kappa}^{B_n},
  \label{eq:ResidualLocalRegion}
\end{equation}
and the new trial- and solution-vector are obtained as in the global Davidson procedure.
The summary of the Davidson algorithm in the local region is given in Algorithm~\ref{alg:davidson_local}.

\begin{algorithm}[htbp!]
  \caption{Davidson algorithm in local region.}
  \label{alg:davidson_local}
  \begin{algorithmic}
    \State \textbf{Compute} $\mathbf{B}_{n_\mathrm{ini}}$, $\bm{\Sigma}_{n_\mathrm{ini}}$, $\mathbf{g}^{B_{n_\mathrm{ini}}}$, $\mathbf{H}^{B_{n_\mathrm{ini}}}$, $\bm{\Lambda}$
    \For{n:=$n_\mathrm{ini}$,$n_\mathrm{ini}+1$, $\dots$ }
      \State \textbf{Solve:} $\bm{\kappa}^{B_n} \gets$ Eq.~(\ref{eq:SubspaceNewton})
      \State $\bm{\kappa}_n \gets  \mathbf{B}_n \bm{\kappa}^{B_n}$
      \State $ \mathbf{r}_n \gets \mathbf{g} + \bm{\Sigma}_n \bm{\kappa}^{B_n}$
      \If{$\lVert\mathbf{r}_n\rVert_2\leq \mathrm{threshold}$}
        \State exit for loop.
      \EndIf
      \State $\mathbf{b}_{n+1} \gets \bm{\Lambda}^{-1}\mathbf{r}_n$
      \State \textbf{Update:} $\mathbf{B}_{n+1}$, $\bm{\Sigma}_{n +1}$, $\mathbf{A}^{B^A_{n+1}}$
    \EndFor
  \end{algorithmic}
\end{algorithm}
/
\subsection*{Implementation of the trust-region optimization strategy}

The iterative optimization is implemented as follows: an initial trust radius $h$ is chosen and the gradient is evaluated.
If the gradient norm is larger than the local region threshold, the global-region Davidson is employed to optimize the orbital rotation matrix.
Otherwise, the local-region one is chosen.
To accelerate convergence we choose the matrix $\mathbf{U}$ in the definition of $\bm{\Theta}=\mathbf{C}\mathbf{U}$ such that the virtual-virtual and occupied-occupied blocks of the Fock matrices are diagonal.\cite{Helmich2021_OrcaTRAH} 
Then, the Davidson algorithm is applied until the norm of the residual falls below the product of the gradient norm and the Davidson threshold scaling.
This dynamic threshold selection avoids unnecessary iterations at the beginning of the optimization.
At convergence, we rotate the molecular orbitals with Eq.~(\ref{eq:Qfactor}) according to
\begin{equation}
  \mathbf{C}_{i+1} = \mathbf{Q}_\mathrm{AO}\mathbf{C}_i,
  \label{eq:MolecularOrbitalRotation}
\end{equation}
after $\mathbf{Q}_\mathrm{AO}$ is Gram--Schmidt orthonormalized.
Subsequently, we update the Fock matrices and check the convergence criteria. 
If the optimization is not converged, the energy change is predicted based on the quadratic model 
\begin{equation}
  \Delta E_\text{pred} = \bm{\kappa}^\mathrm{T} \mathbf{g} + \frac{1}{2} \bm{\kappa}^\mathrm{T} \mathbf{H}(\bm{\kappa}),
  \label{eq:EnergyChangePrediction}
\end{equation}
as well as the actual energy difference between the current and last iterations 
\begin{equation}
  \Delta E_\text{act} = E_{n}-E_{n-1} \, ,
  \label{eq:ActualEnergyChange}
\end{equation}
such that their ratio can be evaluated
\begin{equation}
  r = \frac{\Delta E_\text{act}}{\Delta E_\text{pred}} \, .
  \label{eq:RatioEnergyChange}
\end{equation}

If $\Delta E_\text{act}>0$, the step will be rejected and the trust radius will be decreased as $h = 0.7~h$.
Otherwise, the step is accepted and Fletchers' algorithm \cite{Fletcher2013_PracticalOptimization} is chosen to update the step length according to:

\begin{enumerate}
  \item $0<r<0.25$: he trust radius is decreased to $h=0.7~h$.
  \item $0.25<r<0.75$: the trust radius is left unchanged.
  \item $r>0.75$: the trust radius is increased to $h=1.2~h$.
\end{enumerate}

Special care must be taken if a step in the local region is rejected.
In this case, we first set the trust radius to the current gradient norm.
We then decrease it by a factor of $0.7$, and, at the subsequent iteration, we enforce the trust-radius constraint with the global-region Davidson algorithm.
With the local region threshold of $0.001$, rejecting a step in the local region may occur with ARH method but it is usually not the case with the Newton method.

We provide the default parameters of our implementation in Table~\ref{tab:params}.
``Maximum ARH dimension'' refers to the number of density matrices from previous iterations that are stored. The other parameters are explained in the text.

\begin{table}[htbp!]
  \caption{Default settings for the trust-region augmented Hessian algorithm.}
  \label{tab:params}
  \centering 
  \begin{tabular}{l l} 
    \hline\hline
            Setting               &     default     \\ 
    \hline
    Initial trust radius          &       0.5       \\
    Maximum trust radius          &       1.0       \\
    Maximum Davidson iterations   &       32        \\
    Minimum Davidson threshold    &       0.01      \\
    Davidson threshold scaling    &       0.1       \\
    Local region threshold        &       0.001     \\
    Minimum alpha                 &       1         \\
    Number of start vectors       &       3         \\
    Maximum ARH dimension         &       25        \\
    \hline\hline
  \end{tabular} 
\end{table}


\end{document}